%% file: review.tex
\documentclass[3p,review]{elsarticle}

\usepackage{lineno,hyperref}
\usepackage{color}
\usepackage{mathrsfs}
\usepackage{relsize,amsmath}
\usepackage{amsfonts}
\usepackage{mathtools}
\usepackage{enumitem}
\modulolinenumbers[5]

\journal{Journal of Hydrodynamics}

\begin{document}

\begin{frontmatter}

\title{Spectral/{\it hp} element methods: recent developments, applications, and perspectives}

\author[ic]{Hui Xu\corref{mycorrespondingauthor}}
\cortext[mycorrespondingauthor]{Corresponding author: hui.xu@imperial.ac.uk}
\author[ic]{Chris D. Cantwell}
\author[denmark]{Carlos Monteserin}
\author[aau,rise]{Claes Eskilsson}
\author[denmark,denmark2]{Allan P. Engsig-Karup}
\author[ic]{Spencer J. Sherwin}
\address[ic]{Department of Aeronautics, Imperial College London, London SW7 2AZ, United Kingdom}
\address[denmark]{Department of Applied Mathematics and Computer Science, Technical University of Denmark, 2800 Kgs. Lyngby, Denmark}
\address[denmark2]{Center for Energy Resources Engineering (CERE), Technical University of Denmark, 2800 Kgs. Lyngby, Denmark}
\address[aau]{Department of Civil Engineering, Aalborg University, DK-9220 Aalborg {\O}, Denmark}
\address[rise]{Division Safety and Transport, Research Institutes of Sweden (RISE), SE-50115 Bor{\aa}s, Sweden}

%
%

\begin{abstract}
The spectral/{\it hp} element method combines the geometric flexibility 
of the classical {\it h}-type finite element technique with the desirable 
numerical properties of spectral methods, employing high-degree piecewise 
polynomial basis functions on coarse finite element-type meshes. The spatial 
approximation is based upon orthogonal polynomials, such as Legendre or 
Chebychev polynomials, modified to accommodate $\mathcal{C}^0$-continuous expansions. 
Computationally and theoretically, by increasing the polynomial order {\it p}, 
high-precision solutions and fast convergence can be obtained and, in 
particular, under certain regularity assumptions an exponential reduction in 
approximation error 
between numerical and exact solutions can be achieved. This method has now been 
applied in many simulation studies of both fundamental and practical 
engineering flows. This paper briefly describes the formulation of the spectral/{\it hp} element method and 
provides an overview of its application to computational fluid dynamics. In 
particular, it focuses on the use the spectral/{\it hp} element method in 
transitional flows and ocean engineering. 
Finally, some of the major challenges to be overcome in order to use the
spectral/{\it hp} element method in more complex science and engineering applications are discussed.
\end{abstract}


\end{frontmatter}


\section*{Project Leader}
Spencer Sherwin is Professor of Computational Fluid Mechanics and the Head of 
Aerodynamics Section in the Department of Aeronautics at Imperial College 
London. He received his MSE and PhD from the Department of Mechanical and 
Aerospace Engineering Department at Princeton University in 1995. Prior to this 
he received his BEng from the Department of Aeronautics at Imperial College 
London in 1990. In 1995 he joined the Department of Aeronautics at Imperial 
College as a lecturer and subsequently became a full professor in 2005. Over 
the past 27 years he has specialised in the development and application of 
advanced parallel spectral/{\it hp} element methods for flow around complex 
geometries with a particular emphasis on vortical and bluff body flows, 
biomedical modelling of the cardiovascular system and more recently in 
industrial practice through partnerships with McLaren Racing and Rolls 
Royce. Professor Sherwin’s research group also develops and distributes the 
open-source spectral/{\it hp} element package {\it Nektar}++ (www.nektar.info) 
which has been applied to direct numerical simulation, large eddy simulation 
and stability analysis to a range of applications including vortex flows of 
relevance to offshore engineering and vehicle aerodynamics as well as 
biomedical flows associated with arterial atherosclerosis. He has published 
numerous peer-reviewed papers in international journals covering topics from 
numerical analysis to applied and fundamental fluid mechanics and co-authored a 
highly cited book on the spectral/{\it hp} element method.  Since 2014 Professor 
Sherwin has served as an associate editor of the Journal of Fluid Mechanics. He 
is a Fellow of the Royal Aeronautical Society, a Fellow of the American 
Physical Society and in 2017 he was elected a Fellow of the Royal Academy of 
Engineering.

\input{sec1.tex}

\input{sec2.tex}

\input{sec3.tex}

\input{sec4.tex}

\input{sec5.tex}

\input{acknowledgments.tex}



\section*{References}
\bibliographystyle{mybst}
\bibliography{mybibfile,marineRef}

\end{document}

%% file: sec1.tex
\section{Introduction}
 Over the past few decades, computational fluid dynamics (CFD) has
 become increasingly powerful and has therefore been seen as the
 natural starting point to investigate a variety of mathematical and
 physical problems in science and engineering. Traditionally,
 ``low-order methods'' with up to second-order spatial accuracy have
 been widely adopted as the default implementation for the simulation
 of fluid flows, often based around the Reynolds Averaged Navier-Stokes (RANS)
 equations. This approach has achieved a great deal of success in many
 applications due to its well-established robustness and
 efficiency.  However, as the demands on the accuracy of CFD outputs
 have increased, such as requiring the solution
 of the unsteady flow equations in complex geometries, low-order methods are less able to
 provide the necessary level of precision in capturing transient dynamics, as 
 compared to higher accuracy schemes for a given
 computational cost. Therefore there is currently a great interest in
 the development and application of high-order methods such as the
 spectral/{\it hp} element discretisation.

Finite element methods are widely used across a broad range of
engineering and scientific disciplines. They can be categorised into three 
classes
\cite{BABUSKA19905,Babuska:1994:PHV:209190.209195}: (1) the classical
     {\it h}-version finite element method; (2) the spectral element
     method, or {\it p}-version finite element method; (3) the {\it
       hp}-version finite element method, called the spectral/{\it hp}
     element method.  Once the computational domain is partitioned
     into a non-overlapping element set, the spectral/{\it hp} element
     method employs a ``spectral-like'' approach in each
     element, representing the solution through a basis of polynomials. 
     Therefore, the spectral/{\it hp} element method combines
     the advantages of the spectral element method, in terms of the
     properties of accuracy and rapid convergence, with those of the
     classical {\it h}-version finite element method, that allows
     complex geometries to be effectively captured. Compared with traditional
     low-order finite element schemes, the spectral/{\it hp}
     element method can provide an arbitrary order of
     spatial-approximation accuracy under the assumption of sufficient
     smoothness of the solution. It therefore
     combines the advantages of the low-order finite element method family, 
     whilst
     also providing an additional attractive higher-precision,
     approximation for solving partial differential equations
     \cite{spectralhp2005,CANTWELL2015205}.
Recent studies also indicate that the compact nature of the
spectral/{\it hp} element method is well-placed to take advantage of
modern multi-core and many-core computing hardware
\cite{Yakovlev2016,King2014}. All of the above properties have
positioned the spectral/{\it hp} element method as an attractive numerical
strategy for obtaining high-precision solutions with a relatively low
computational cost. 

The spectral/{\it hp} element method is gaining increasing traction
in the field of CFD \cite{spectralhp2005}, and it has achieved great success 
in both the direct and large eddy simulation (LES) of complex flows
\cite{lombard2016,SERSON2016243,SERSON2017117,xu_lombard_sherwin_2017,xu_mughal_gowree_atkin_sherwin_2017}.
It has also been successfully applied to a broad range of other applications in 
various research fields, for instance, cardiac electrophysiology
\cite{CANTWELL2014813}, solid mechanics \cite{szabo91,MOXEY2014127}, porous 
media \cite{Comerford2015} and oceanographic modelling \cite{ESKILSSON2006566}.


One of the challenges of these methods is that they are more challenging
to implement than low-order methods. There are now, however, a number of open-source packages which encapsulate the complexities of the method. One such package that the authors have been
developing is {\it Nektar}++, a cross-platform spectral/{\it hp}
element framework (\url{http://www.nektar.info}) which has made high-order
finite element methods more accessible to the broader community and can be used 
to solve a range of the emerging challenges in high-fidelity scientific 
computations \cite{CANTWELL2015205}. It enables the construction of high-order discretisations for solving a wide range of partial differential equations, supporting hybrid meshes, for example, triangles and quadrilaterals or tetrahedra, prisms, pyramids and hexa-hedra. The current software package includes a number of pre-written scalable solvers, including for incompressible flow, compressible flow, shallow water equations, acoustic perturbation equations and others.

In this paper, we present a review of the state-of-the-art of the
spectral/{\it hp} element method, and its applications.
In \S 2, we briefly describe
the spectral/{\it hp} element method and discuss the numerical
advantages and robustness of the method in
solving partial differential equations. In \S 3, we highlight some of the
implementational challenges of the method in order to achieve computational 
efficiency and discuss potential solutions through our efforts with {\it 
Nektar}++. In \S 4, we then provide a survey of the applications in 
transitional flows and waves  (wave propagation and wave-body interaction) in ocean engineering to which the spectral/{\it hp} element 
method has been applied. Finally, in \S 5, we conclude with further discussions 
and some perspectives on future directions.

%% file: sec2.tex
\section{Spectral/{\it hp} element method}

In this section we provide an overview of the mathematical foundations of the
spectral/{\it hp} element method. A more detailed derivation of the 
mathematical theory can be found in \cite{spectralhp2005} but is beyond
the scope of this paper.



\begin{figure}[h!]
  \centering
    \includegraphics[width=0.4\columnwidth]{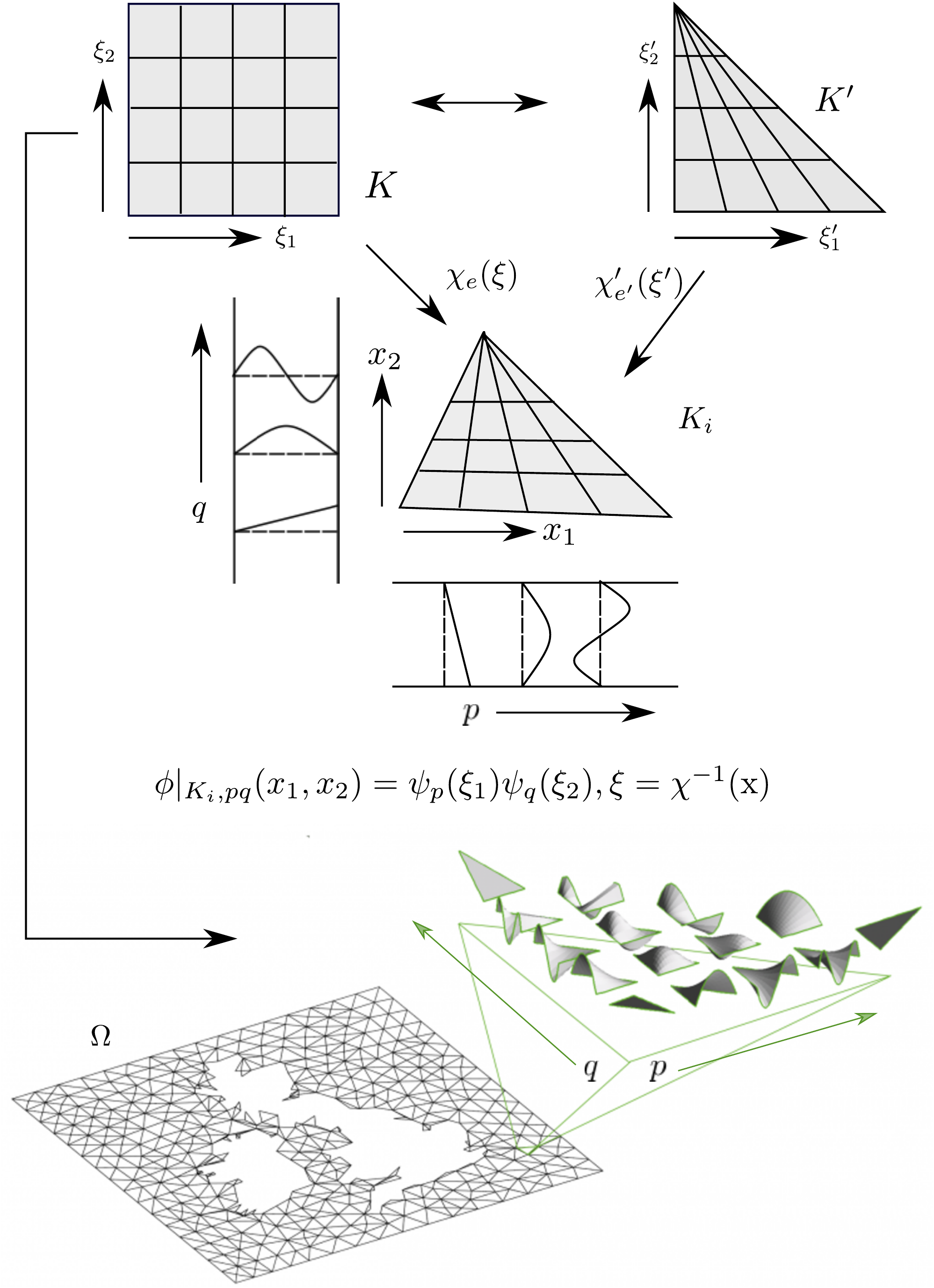}
\caption{Illustration of the construction of a two-dimensional
  fourth-order $\mathcal{C}^0$-continuous modal triangular expansion
  basis using a generalised tensor-product procedure.}
  \label{fig2}
\end{figure}

\subsection{Overview of method}
Spectral/{\it hp} element discretisations are the underpinning
approximation for both continuous and discontinuous Galerkin
formulations and can be constructed in 1-D, 2-D and 3-D. To obtain a general view of the method we can consider the example in Fig. \ref{fig2}. At the bottom of this figure we observe a
triangulation of the British Isles. In general, the computational mesh can 
comprise of a mix of
different shaped elements which could be triangles and quadrilaterals
in two dimensions, and tetrahedra, pyramids, prisms and hexahedra in
three dimensions. Within each element we develop a polynomial
expansion, as represented in the lower-right of Fig.~\ref{fig2}, where we
observe all of the expansion modes to represent a fourth-order
polynomial using a modal or hierarchical expansion basis. For
Continuous Galerkin expansions, the design of these elements typically has
the property of being decomposed into boundary and interior modes  so that
$\mathcal{C}^0$-continuity can be achieved by matching boundary expansion modes
that have support on the edges of the element. The interior modes are
necessarily defined to be zero on the boundaries.

There are a number of ways of developing the polynomial expansion
within an element $K_i$. In the example shown in Figure~\ref{fig2} we
illustrate the popular construction using a tensor product of
one-dimensional expansions which are defined in the regular regions
$K$ and $K'$ which can then be mapped to the physical elemental regions using
the mapping $\vec{\chi}(\vec{\xi})$. Tensor-based expansions are very
popular in quadrilateral and hexahedral regions but require some
modification for triangular or tetrahedral regions using a Duffy 
\cite{duffy1982} or
collapsed-coordinate system (see \cite{spectralhp2005} for more details). It is
possible to define polynomial expansions that are hierarchical/modal
in construction, much like a small-scale spectral expansion, or alternatively 
one
can use a nodal basis where the expansion is defined at a specific set of points
(often related to Gaussian quadrature) and only one expansion mode has a
unit value, whilst all other expansions basis are zero, at this
point. This provides a Kronecker delta property to the expansion which
can be useful when evaluating non-linear products. Again, for
triangular elements the construction of a nodal basis is more involved
and typically requires the use of a generalised Vandemonde matrix to
relate them to a known expansion which is often of a modal
construction (see \cite{hesthaven}).

Once a polynomial expansion has been defined in each element,
the approach used to ``bolt together'' these individual expansions
defines the approximation
method. For example, if one enforces a $\mathcal{C}^0$-continuous expansion by
ensuring the polynomial expansion is continuous across elemental
boundaries we obtain the classical continuous Galerkin
method. Alternatively, if one does not directly enforce the expansions
to be $\mathcal{C}^0$-continuous but ensures appropriate flux quantities are
continuous between elements, one can construct a so-called
Discontinuous Galerkin scheme \cite{spectralhp2005,hesthaven,hill1973}.

\subsection{Outline of mathematical formulation}
In general we consider the numerical solution of partial differential
equations (PDEs) of the form $\mathcal{L}u = 0$ on a domain $\Omega$, which may be
geometrically complex, for some solution $u$. Practically, $\Omega$
takes the form of a $d$-dimensional finite element mesh consisting of elements 
$K_i$, embedded in a space of dimension $\hat{d}$, such that $d \leq \hat{d} 
\leq 3$, with $\Omega = \cup_i K_i$ and
$K_i\cap K_j$ is an empty set or an interface between elements of dimension
$\bar{d} < d$. We solve the PDE problem in the weak sense and, in
general, $u|_{K_i}$ must be smooth with at least a first-order
derivative; we therefore require that $u|_{K_i}$ is in the Sobolev space
$W^{1,2}(K_i) \equiv H^1(K_i)$ \cite{adams}. For a continuous discretisation, 
we additionally
impose continuity along element interfaces.  For illustrative purposes, we 
assume that it can be expressed as follows:
\begin{equation}\label{op}
\mbox{find $u\in H^1(\Omega)$  such that }a(u,v)=l(v)\ \forall v \in H^1(\Omega).
\end{equation}
where $a(\cdot,\cdot)$ is a bilinear form, $l(v)$ is a linear form and $H^1(\Omega)$ is formally defined as
\begin{equation}
  H^1(\Omega) :=\{v\in L^2(\Omega)|D^\alpha u\in L^2(\Omega)\ \forall\ |\alpha|\leq 1\}.
  \end{equation}
To solve this problem numerically, we consider solutions in a
finite-dimensional subspace $V^\delta \in H^1(\Omega)$ and cast our problem as:

\begin{equation}\label{op1}
\mbox{find $u^\delta \in H^1(\Omega)$  such that 
}a(u^\delta,v^\delta)=l(v^\delta)\  \forall\ v^\delta \in V^\delta,
\end{equation}
augmented with appropriate boundary conditions. For a projection which
enforces continuity across elements, we impose the additional
constraint that $V^\delta \in \mathcal{C}^0$. We assume the solution
can be represented as $u^\delta({\bf x}) = \sum_n \hat{u}_n
\Phi_n({\bf x})$, a weighted sum of $N$ trial functions $\Phi_n({\bf
  x})$ defined on $\Omega$ and our problem becomes that of finding the
coefficients $\hat{u}_n$. The approximation $u^\delta$ does not
directly give rise to unique choices for the coefficients
$\hat{u}_n$. To achieve this we place a restriction on the residual $R = 
{\mathcal
  L}u^\delta$ that its $L^2$ inner product, with respect to the test
functions $\Psi_n({\bf x})$, is zero. For a Galerkin projection we
choose the test functions to be the same as the trial functions, that
is $\Psi_n = \Phi_n$.  As outlined previously, to construct the global
basis $\Phi_n$ we first consider the contributions from each element
in the domain. Each $K_i$ is mapped from a standard reference space $K
\in [−1, 1]^d$ by a parametric mapping ${\chi}_e : K\rightarrow K_i$
given by ${\rm x} = {\chi}_e({\xi})$, where $K$ is one of the supported
region shapes, and ${\xi}$ are $d$-dimensional coordinates representing
positions in a reference element, distinguishing them from ${\rm x}$ which
are $\hat{d}$-dimensional coordinates in the Cartesian coordinate
space. The mapping need not necessarily exhibit a constant Jacobian,
supporting deformed and curved elements through an isoparametric
mapping. For triangular, tetrahedral, prismatic and pyramid elements, the Duffy 
transform \cite{duffy1982} is used and operations, such as
calculating derivatives, map the coordinate system to the non-collapsed
coordinate system.

A local polynomial basis is constructed on each reference element with
which to represent solutions. A one-dimensional order-$P$ basis is a set
of polynomials $\phi_p(\xi), 0 \leq p \leq P$, defined on the
reference segment. The particular choice of basis is usually made based on
its mathematical or numerical properties and may be modal or nodal in
nature. For two- and three-dimensional regions, a tensorial basis may
be used, where the polynomial space is constructed as the
tensor-product of one-dimensional bases on segments, quadrilaterals
or hexahedral regions. In particular, a common choice is to use a
modified hierarchical Legendre basis, given as a function of one
variable by
\begin{equation}
\phi_p(\xi)\left\{
\begin{array}{ll}
\frac{1-\xi}{2}, & p=0,\\
\left(\frac{1-\xi}{2}\right)\left(\frac{1+\xi}{2}\right)P^{1,1}_{p-1}(\xi), & 0<p<P,\\
\frac{1+\xi}{2}, & p=P,
\end{array}
\right.
\end{equation}
which supports the aforementioned boundary-interior decomposition and therefore 
improves numerical efficiency when solving the globally assembled
system. Equivalently, $\phi_p$ could be defined by the Lagrange polynomials
through the Gauss-Lobatto-Legendre quadrature points which would lead
to a traditional spectral element method.  On a physical element $K_i$
the discrete approximation $u^\delta$ to the solution $u$ may be expressed as
\begin{equation}
 u^\delta({\rm x})=\sum_{n}\hat{u}_n\phi_n([{\chi}_e]^{-1}({\rm x}))
\end{equation}
where $\hat{u}_n$ are the coefficients from
Eq. (\ref{op}), obtained through projecting $u$ onto the discrete
space. Therefore, we restrict our solution space to
\begin{equation}
V\coloneqq\{u\in H^1(\Omega)\ |\ u|_{K_i}\in \mathcal{P}_p(K_i)\}
\end{equation}
 where $\mathcal{P}_p(K_i)$ is the space of order-$p$ polynomials on $K_i$. Elemental
 contributions to the solution may be assembled to form a global
 solution through an assembly operator. In a continuous Galerkin
 setting, the assembly operator sums contributions from neighbouring
 elements to enforce the $\mathcal{C}^0$-continuity requirement. In a discontinuous
 Galerkin formulation, such mappings transfer flux values from the
 element interfaces into global solution vector.

\begin{figure}[h!]
  \centering
    \includegraphics[width=0.9\columnwidth]{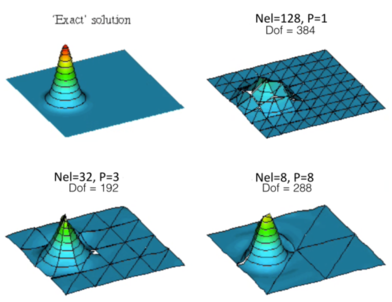}
\caption{Propogation of a cone initial condition using a DG formulation of the advection equation. After one rotation we see the solution using a $P$=2, $P$=6 and $P$=8 order approximations. }
  \label{cone}
\end{figure}

\subsection{What are the beneficial properties of using a spectral/{\it hp} element method?}

For smooth problems one of the benefits of a spectral/{\it hp}
formulation is often quoted as the exponential rate of convergence one
observes when increasing the polynomial order of the approximation,
providing the solution has sufficient regularity. In practice, the good
approximation properties are also realised in terms of the better
diffusion and dispersion properties one observes as you increase the
polynomial order. This is exemplified in Fig. \ref{cone} where we have
taken an initial condition of a Gaussian and advected it under a rotational 
velocity field for one revolution. We then observe the solution of a
DG approximation of the linear advection problem using a $P$=1, $P$=3 and
$P$=8 order approximation with $128,32$ and $8$ elements, respectively. We see from
this simple test that the shape is far less diffused in the $P$=8
problem when compared to the $P$=2 discretisation, even though the overall 
number of degrees of freedom is comparable.

Recently, a comparison between high-order flux reconstruction and the 
industrial-standard solver STAR-CCM+ was made, in which the accuracy and 
computational cost of flux reconstruction (FR) methods in PyFR and STAR-CCM+ 
for a range of test cases including scale-resolving simulations of turbulent 
flow  were investigated \cite{VERMEIRE2017497}.  The results from both PyFR and 
STAR-CCM+ show that third-order schemes can be more accurate than second-order 
schemes for a given cost. Moreover, advancing to higher-order schemes on GPUs 
with PyFR was found to offer even further accuracy vs cost benefits relative to 
industry-standard tools. These  demonstrate the potential utility of high-order 
methods on GPUs for scale-resolving simulations of unsteady turbulent flows.
  
\subsection{Using the spectral/{\it hp} element method in marginally resolved problems}

In the previous sections we outlined the underlying
formulation of the spectral/{\it hp} element methodology and we now describe
some of the state-of-the-art developments necessary for simulating problems 
which are only marginally resolved.  These tend to arise
at higher Reynolds numbers, which generally promote greater levels
of turbulence that cannot realistically be captured, even on the latest
parallel computers. This leads to the build-up of energy in the larger resolved 
scales, due to the inability of the discretisation to fully capture the smaller 
scales of the flow. This is addressed with two techniques. The first involves 
performing the numerical integration of non-linear terms consistently. Even 
with this, non-linear interactions may still lead to an artificial build-up of 
energy in the smaller resolved scales. A second technique is then to apply a 
spectral filtering, known as spectral vanishing viscosity, to dampen such 
energies.

\subsubsection{Dealiasing}

Dealiasing strategies have been observed to be effective in enhancing
the numerical stability when solving problems using the  spectral/{\it hp} element
method \cite{gottlieb_orszag, KIRBY2003249,Kirby2006,
  malm_schlatter_henningson_2012, MENGALDO201556}. The errors are
typically caused by {\it insufficient} quadrature employed in the
Galerkin discretisation of nonlinear terms \cite{spectralhp2005}. 
When the simulation is not sufficiently well-resolved, this leads to energy
in the shorter length scales to be transferred back into the longer
length scales. When simulations are well-resolved, the numerical errors created 
by insufficient quadrature are negligible
\cite{KIRBY2003249,Kirby2006}. However, in marginally resolved or
under-resolved simulations, aliasing errors may significantly pollute
the solution \cite{KIRBY2003249}. As discussed in
\cite{MENGALDO201556}, there are three kinds of aliasing
sources:
\begin{itemize}
\item PDE aliasing, which relates to quasi-linear and non-linear
  terms (see \cite{KIRBY2003249} for a simple example of
  the Burgers equation);
\item geometrical aliasing, which arises due to deformed/curved elements.
  \item interface-flux aliasing, in the case of discontinuous methods (see
    \cite{MENGALDO201556} for more details).
\end{itemize}

\begin{figure}[h!]
  \centering
\includegraphics[height=0.2\columnwidth]{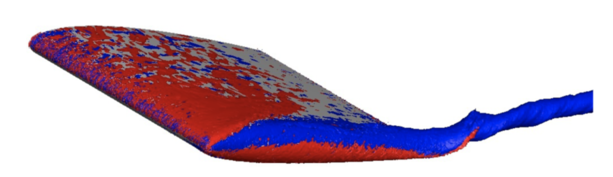}
\caption{Helicity for an incompressible viscous flow over a NACA 0012 wing (from the reference \cite{MENGALDO201556})}
  \label{naca0012a}
\end{figure}

\begin{figure}[h!]
  \centering
\includegraphics[height=0.17\columnwidth]{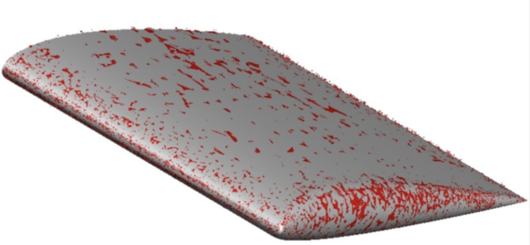}
\includegraphics[height=0.17\columnwidth]{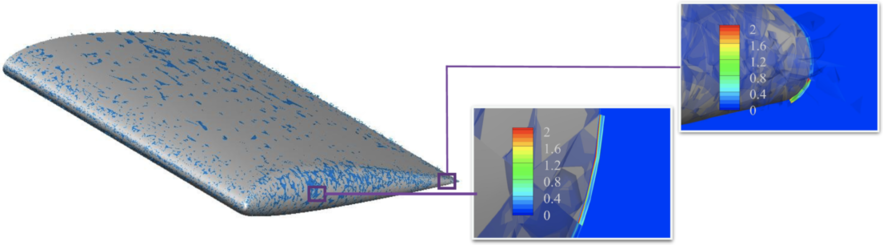}\\
\caption{PDE aliasing errors at 30\% of the magnitude of the nonlinear terms 
(a) and close up regions near the wing surface showing geometric aliasing 
errors (b) (from the reference \cite{MENGALDO201556})}
\label{naca0012b}
\end{figure}

We start by giving an example from incompressible viscous flow past a
NACA 0012 wing tip at a Reynolds number of $Re=1.2\times10^6$, originally 
discussed in \cite{MENGALDO201556}, to
demonstrate the aliasing error in the near surface region.
Figure~\ref{naca0012a} illustrates the dynamics of the flow by showing
the iso-contours of helicity with a $P$=3 approximation. In
the boundary region, prismatic elements are employed while tetrahedral
elements are used in the rest of the domain. Figure~\ref{naca0012b}, shows the
first two kinds of aliasing errors. In Figure~\ref{naca0012b}(a), 30\%
aliasing error with respect to the magnitude of the nonlinear terms is
observed and in Figure~\ref{naca0012b}(b) we show a close up of regions with 
high geometrical aliasing error near the wing surface where 600\% error is 
observed. The aliasing errors in this numerical calculation 
significantly pollute the solution and cause it to very rapidly
become numerically unstable.


We can categorise dealiasing techniques into two strategies. 
\begin{itemize}
\item {\it Local dealiasing}: PDE-dealiasing through consistent
  integration of the nonlinear terms only
  \cite{KIRBY2003249,MENGALDO201556};
\item {\it Global dealiasing}: PDE- and geometrical-dealiasing through
  consistent integration of all the terms of the discretisation
  \cite{MENGALDO201556}.
\end{itemize}
The phrase {\it consistent integration} refers to polynomial
nonlinearities, where it is effectively possible to know a priori the
number of quadrature points necessary for the integration to be exact
\cite{KIRBY2003249,cockburn1990}. Therefore, the non-linearities can be
consistently integrated. For non-polynomial functions, as typically
arise in the Euler equations, the concept of consistent integration is
not well defined since it is impossible to fully control the
quadrature error. We refer to the reader to \cite{MENGALDO201556} for
a more complete discussion of these strategies as well as a discussion
of interface aliasing.

\subsubsection{Spectral vanishing viscosity}\label{svvsec}

As previously discussed, it is well known that spectral/{\it hp}
element discretisations generally lead to approximations that have low
dissipation and low dispersion \cite{spectralhp2005} per degree of
freedom when compared to lower-order methods. In solving
advection-diffusion equations and nonlinear partial differential
equations such as advection-dominated flows, at marginal resolutions,
oscillations appear that may render the computation unstable
\cite{spectralhp2005}. Artificial viscosity has been used in may
discretisation methods to suppress wiggles associated with high
wavenumbers, for example hyperviscosity has been broadly and
effectively used in simulations using the Fourier method.  A related
concept is the so-called {\it spectral vanishing viscosity} (SVV),
which was originally proposed based on a second-order diffusion
(convolution) operator for spectral Fourier methods \cite{tadmor}. The
SVV concept was originally motivated through the inviscid Burgers
equation where an additional diffusion term was added, {\it i.e.} 
\begin{equation}\label{vburgers}
\frac{\partial }{\partial t}u(x,t)+\frac{\partial}{\partial 
x}\left(\frac{u^2(x,t)}{2}\right)=\epsilon\frac{\partial}{\partial 
x}\left(\mathscr{D}_\epsilon\frac{\partial u(x,t)}{\partial x}\right),
\end{equation}
where $\epsilon(\rightarrow 0)$ is a viscosity amplitude and
$\mathscr{D}_\epsilon$ is a viscosity kernel, which may be nonlinear,
can be a function of $x$ and is only activated for high wave
numbers. With a small amount of controlled dissipation, spectral
accuracy can be retained. This approach was extended to the
Navier-Stokes equations by Karamanos \& Karniadakis
\cite{KARAMANOS200022} and Kirby \& Sherwin \cite{KIRBY20063128} and
has also been investigated and applied to Large Eddy Simulation by
Pasquetti \cite{pasquetti2005,Pasquetti2006}.

\begin{figure}[h!]
  \centering
    \includegraphics[width=0.8\columnwidth]{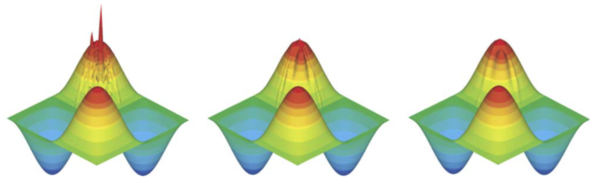}
\caption{A demonstration of stabilisation by SVV: (left) standard
  diffusion at $t$=0.1; (centre) standard diffusion with SVV $P_{\rm
    cut}$=7, $\epsilon$=0.1; (right) standard diffusion with SVV
  $P_{\rm cut}$=3, $\epsilon$=0.1 ( from the reference \cite{KIRBY20063128}).}
  \label{svv}
\end{figure}

To provide some intuition about the influence of SVV, we show in 
Figure~\ref{svv} an
example from \cite{KIRBY20063128} of applying SVV to the
parabolic equation
\begin{equation}
\frac{\partial u}{\partial t}=\nu\nabla^2 u+S_{VV}(u)
\end{equation}
on $\Omega=[0,2]\times[0,2]$ with $\nu=10^{-5}$ and periodic boundary
conditions. The initial condition is given by $u(x,y,t=0)=\sin(\pi
x)\sin(\pi y)$, in which a perturbation is introduced. A total of
sixteen quadrilateral elements (four per direction) with $15th$ order
polynomials per element direction were used. The following ``classical''
one-dimensional SVV kernel is employed,
\begin{equation}
\mathscr{D}=\left\{\begin{array}{ll}
0 & p\leq P_{\rm cut},\\
\exp\left(-\frac{(p-P)^2}{(p-P_{\rm cut})^2}\right) & p> P_{\rm cut},
\end{array}
\right.
\end{equation}
where $P$ is the total number of modes employed and $P_{\rm cut}$ is
the cutoff polynomial order. SVV with the kernel function
$\mathscr{D}$ can be regarded as a low-pass filter. We see that the
SVV dissipation added to the high mode numbers with respect to the
spectral element discretisation does indeed yield dissipation at the
global high wavenumber scales of the solution in Fig. \ref{svv}.

In applications, Karamanos and Karniadakis applied SVV to high
Reynolds number turbulent flows and the method was viewed as an {\it
  alternative} LES approach \cite{KARAMANOS200022}. Kirby and
Karniadakis computed the artificial viscosity by incorporating the
local strain and the Panton function and the methodology was called
SVV-LES \cite{kirby2002}. The SVV approach has been reasonably widely
used to simulate turbulent flow and vortical flows
\cite{lombard2016,KIRBY20063128,pasquetti2005,Pasquetti2006,Xu2006,SEVERAC20071234,KOAL20123389}.  SVV has been explicitly regarded as a
turbulent model of implementing implicit large eddy simulation (iLES)
under the assumption ``{\it The action of subgrid scales on the
  resolved scales is equivalent to strictly dissipative action}"
\cite{sagaut}, even though SVV is not explicitly designed as a
subgrid-scale model.

Recently, in order to achieve a better performance of SVV, the linear
advection equation augmented with spectral vanishing viscosity (SVV)
has been analysed \cite{MOURA2016401}. Dispersion and diffusion
characteristics of the spectral/{\it hp} continuous Galerkin (CG)
formulation with SVV-based stabilization are verified to display
similar non-smooth features in flow regions where convection is much
stronger than dissipation or vice-versa, owing to a dependency of the
standard SVV operator on a local P{\'e}clet number $Pe=c h/\epsilon$ where $a$ is the advection speed, $h$ is the mesh spacing and $\epsilon$ is the base SVV magnitude \cite{MOURA2016401}.  A modification
was proposed \cite{MOURA2016401} to the traditional SVV scaling, which 
enforces a globally
constant P{\'e}clet number. In this approach the artificial
dissipation strength parameter $\epsilon$ is defined by
$\epsilon=\epsilon_0 c h/P$, where $\epsilon_0$ is a globally fixed
parameter used to adjust the dissipation strength, $P$ would be the
polynomial order used in each element and $c$ is a measure of the
advection velocity magnitude on each local element.  Furthermore, the
following ``power'' kernel function is proposed \cite{MOURA2016401}
\begin{equation}
\mathscr{D}=(p/P)^{P_{SVV}},
\end{equation}
where $p$ denotes mode index and $P_{SVV}$ is now not an activation
threshold, but a similar effect in terms of computing the viscous
effects on the highest modes. The mesh P{\'e}clet number is now held constant
globally. In addition, the ``power kernel'' function has been devised
for the advocated SVV operator to provide a consistent increase in
resolution power (per degree of freedom) when the polynomial order is
increased -- a feature not naturally achieved through the widely used
“exponential kernel” introduced in \cite{tadmor1993}.

In turbulence simulations using Discontinuous Galerkin methods, there
were early discussions \cite{scollis2002,scollis2003}, highlighting
that the scheme successfully predicted low-order statistics with fewer
degrees of freedom (DoFs) than traditional numerical methods.  More
extensive assessments \cite{WEI201185,CHAPELIER2014210} indicated that
DG can predict high-order statistics with accuracy comparable to that of
spectral methods for an equivalent number of DoFs. However, there is
still little understanding of {\it why} and {\it how} one can use the
spectral/{hp} element method for under-resolved turbulence simulations
when either spectral vanishing viscosity (as sometime employed in
Continuous Galerin methods) or upwind fluxes (as naturally arise in DG
methods) are used to provide dissipation at under-resolved
scales. Recently, exploratory studies, focusing on the DG formulation, were undertaken to address how numerical dissipation affects the under-resolved scales \cite{MOURA2015695}.  By exploring DG's
propagation characteristics directly from the dispersion-diffusion
curves of a linear advection problem, a simple criterion (named `the
1\% rule') was proposed to estimate the effective resolution of the DG
scheme for a given {\it hp} approximation space.

Numerical experiments on Burger turbulence helped to clarify {\it why}
the DG formulation can be suitable for under-resolved simulations of
turbulence: the numerical discretisation is capable of resolving
scales up to $k_{1\%}$ with good accuracy while dissipation is
provided at the end of the energy spectrum in the form of numerical
diffusion.  A further study for improving SVV by the ``1\%'' rule was
conducted with a new SVV kernel function\cite{MOURA2016401}. The
``1\%'' rule enables the SVV to mimic the property of the DG method
with upwind numerical fluxes. A CG formulation based on the suggested
SVV operator and kernel function was investigated.  Being able to
emulate the upwind properties of a DG scheme within a CG approximation
using SVV appears to have relatively robust properties which are
attractive for the marginally resolved simulations and this approach
is referred to as a `DGKernel' SVV within the {\it Nektar}++ framework
discussed in the next section.

%% file: sec3.tex
\section{Implementation of spectral/{\it hp} element method}
One of the challenges of enabling the spectral/{\it hp} element method to become more 
pervasive for industrial and environmental problems is the implementational 
complexity of the methods. One of the overarching aims of the {\it Nektar}++ project 
is to provide an efficient framework upon which a broad range of physical 
processes can be modelled using these approaches. With this in mind, we outline 
some of efficiency challenges associated with implementing these methods for 
practical use and detail some suitable solutions in the context of {\it Nektar}++.

\subsection{Nektar++ overview}
\begin{figure}
\centering
    \includegraphics[width=0.5\columnwidth]{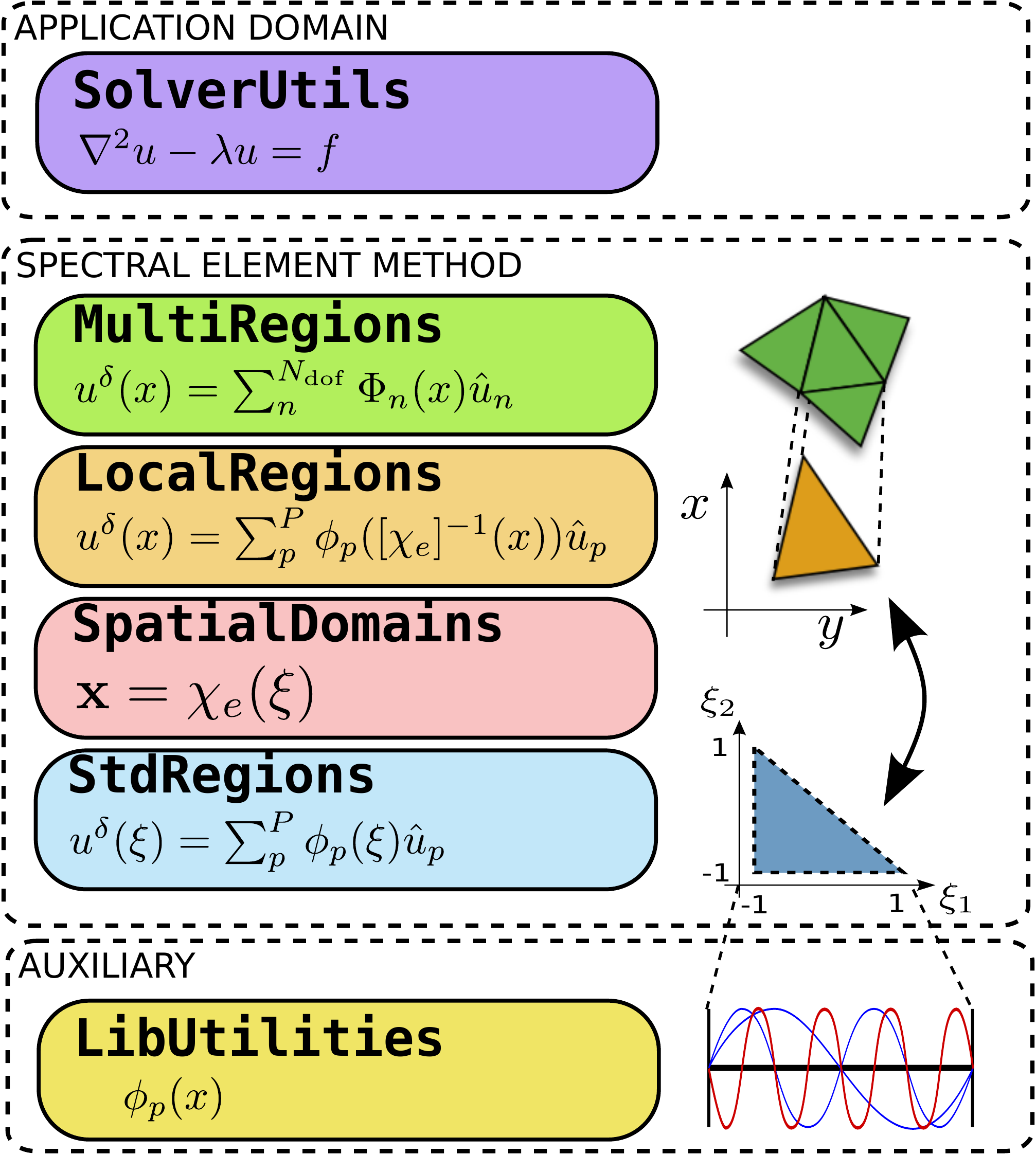}
    \caption{Diagrammatic representation of the libraries in {\it Nektar}++ (from the reference \cite{CANTWELL2015205}).}
    \label{f:nek-impl}
\end{figure}
The design of {\it Nektar}++ is intended to mirror the mathematical formulation of the spectral/hp element method, from one-dimensional basis functions up to multi-dimensional, multi-element  
discretisations of complex geometries. The package consists of a tiered 
collection of libraries which implement different aspects of the formulation, 
as illustrated in Figure~\ref{f:nek-impl}. On top of this, domain-specific 
application solvers can be readily developed which specify the physical 
processes to be modelled, while leveraging the implementation of the 
discretisation provided by the libraries in a relatively transparent manner.

At the lowest level the {\it LibUtilities} implements one-dimensional bases of orthogonal 
polynomials which are used to construct the expansion bases on individual 
elements. Their analytical derivatives and the distributions of points used for 
performing Gaussian quadrature are also captured. On top of this, the reference 
elements are defined for tessellating one-, two- and three-dimensional domains 
in the {\it StdRegions} library. {\it Nektar}++ supports a range of elemental regions to 
enable maximum flexibility in defining meshes which can effectively capture 
complex geometries, including segments in 1D, quadrilaterals and triangles in 
2D, and hexahedra, prisms, pyramids and tetrahedra in 3D. For multi-dimensional 
regions expansion bases may be defined using natively two-dimensional bases 
({\it e.g.} Lagrange polynomials at Fekete points), or though a tensor-product 
construction of 1D polynomials. For each of these regions, core finite element 
operators, including backward transform, inner product and derivative operators 
are defined.

Constituent elements of the computational mesh are represented in the 
{\it LocalRegions} library through a mapping from the reference region to the 
physical coordinates of the element. These mappings are captured in the {\it
SpatialDomains} library and can capture high-order geometric curvature in the 
element. These geometric factors or used to extend the finite element operators 
defined for the standard reference regions to the physical elements. The
{\it MultiRegions} encapsulates the construction of multi-element global expansions 
through the construction of assembly operators which identify corresponding 
degrees of freedom in the local and global representations. Through this 
mapping, global finite element operators are constructed. Global linear systems 
can then be solved using a number of direct and iterative linear algebra 
techniques. Iterative approaches, such as the preconditioned conjugate gradient 
method, solve a system through the repeated action of the operator, providing 
scope for a range of performance optimisations, discussed in Section 3.3.

While the central concept captured in {\it Nektar}++ are high-order spectral/{\it hp} 
element spatial discretisations, high-order time integration algorithms are 
also implemented to allow for highly accurate transient simulations. These 
include fully explicit, implicit and mixed implicit-explicit (IMEX) schemes.

\subsection{Combined spectral/hp-Fourier discretisation}
While the spectral/{\it hp} element method exhibits some of the numerical 
characteristics of a pure spectral method, the use of the latter is still 
preferred in the case of regular geometries such as rectangles and cuboidal 
domains. This situation occurs widely in many, particularly fundamental, 
hydrodynamics problems. In {\it Nektar}++, we leverage this by allowing for a mixed 
spectral/{\it hp}-Fourier discretisation. In this configuration, geometrically 
homogeneous coordinate directions are represented with a truncated Fourier 
expansion, while coordinate directions with geometric complexity use the 
spectral/hp element discretisation. This has the effect of decoupling the 
individual Fourier modes, allowing the spectral/{\it hp} operator to be applied independently 
within each component of each Fourier mode.

The Fourier decomposition also enables parallel decomposition of the modes to 
different processes \cite{BOLIS201617}. Combined with the parallelisation of 
the spectral/{\it hp} 
element discretisation this leads to a hybrid parallelism. An open challenge 
here is identifying the optimal parallelisation of such mixed-discretisation 
domains to achieve optimal performance.

The {\it GlobalMappings} library further extends this discretisation strategy by 
supporting geometries in which a coordinate direction can be transformed to a 
homogeneous representation by an analytical function. This enables more complex 
geometries to be discretised using this efficient formulation.

\subsection{From h to p efficiently}
One of the challenges in implementing the spectral/{\it hp} method is achieving 
computational performance across a broad range of polynomial orders. The 
classic approach to implementing linear finite element operators is to assemble 
the global form of the operator and act directly on the global degrees of 
freedom. This is performant since mesh vertex  has a high valency and 
the assembly operation dramatically reduces the size of the global linear 
system used within the iterative solver. With increasing polynomial order, this 
reduction is less and the memory indirection associated with the global 
operator leads to reduced through-put and degraded performance.

A number of studies \cite{VOS20105161, CANTWELL2011A, CANTWELL2011B, BOLIS2014} 
have concluded that for higher polynomial orders, one 
should instead scatter the global degrees of freedom back onto their local 
elemental representation and perform the action of the operator in an 
element-wise fashion using a matrix-vector operation. This enables a more 
compact representation of the operator in memory, greater cache-locality in 
applying the operator and a reduction in overall memory usage. At very high 
polynomial orders, the size of the local operator grows as $\mathcal{O}(P^4)$. 
A more efficient approach in these cases is to take advantage of the 
tensor-product construction of the operators to decompose the elemental 
operator into a sequence of one-dimensional operators, known as 
sum-factorisation. Algebraically, this 
corresponds to a sequence of matrix-matrix operations which, besides the 
reduction in floating 
operation count, can be generally more efficiently implemented.

Recalling the construction of the local elemental operators as the reference 
region operator, transformed under a geometric mapping, one can exploit the 
decomposition of the elemental operators into these constituent parts. The 
{\it Collections} library in {\it Nektar}++ exploits this decomposition on groups of 
similar elements (same shape and expansion basis) by applying first the 
geometric factors associated with the operator, before acting with the 
reference region operator on all of the elements in the group simultaneously 
\cite{MOXEY2016628}. 
This effectively replaces a sequence of matrix-vector operations with a single 
matrix-matrix operation, reducing data movement and improving floating-point 
performance. This decomposition can be applied both to the local elemental 
matrix representation as well as the sum-factorisation approach.

\subsection{High-order mesh generation}
\begin{figure}[h!]
\centering
    \includegraphics[width=0.6\columnwidth]{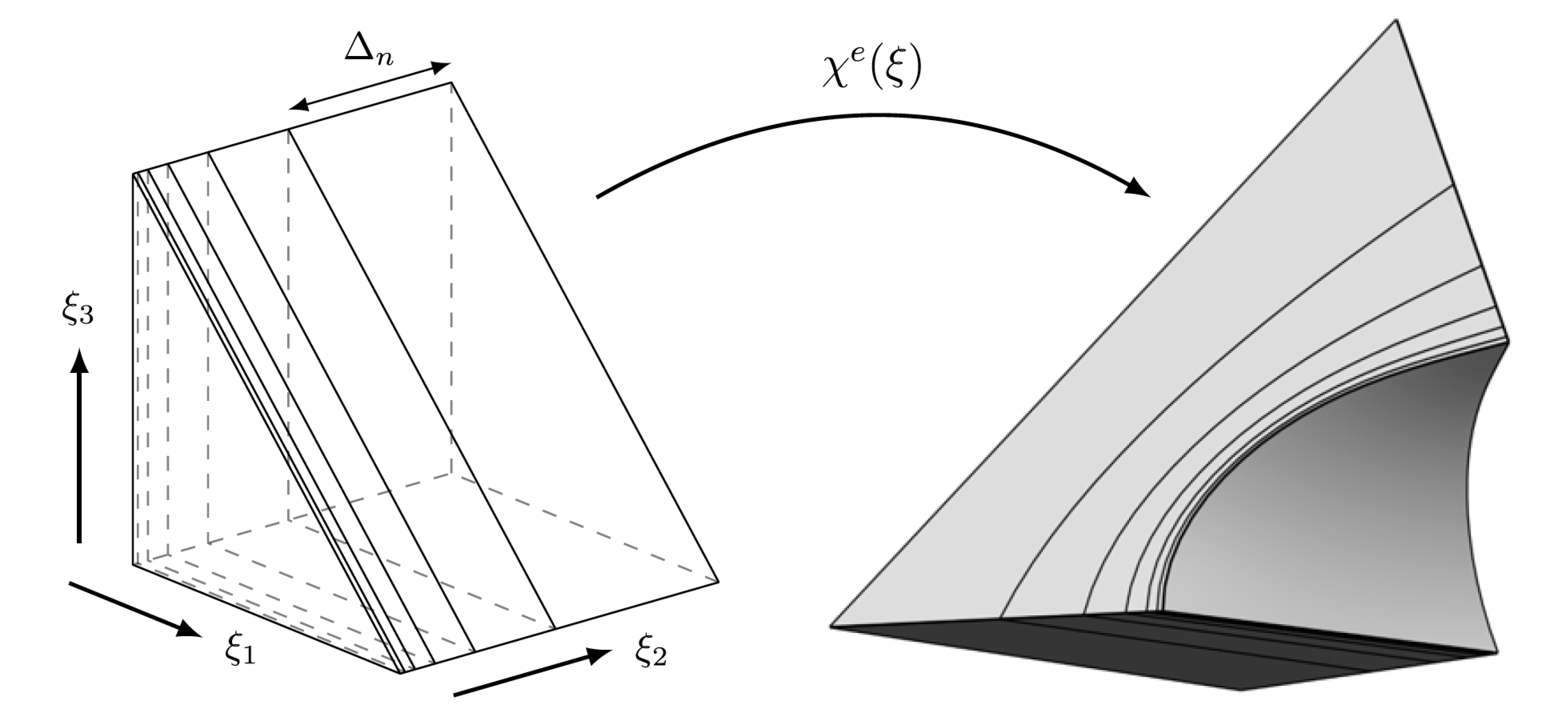}
    \caption{Splitting a reference prismatic element and applying the mapping to obtain a high-order layer of prisms from the macro-element (from the reference \cite{MOXEY2015636})}
    \label{homesh}
\end{figure}
\begin{figure}[h!]
\centering
    \includegraphics[height=0.3\columnwidth]{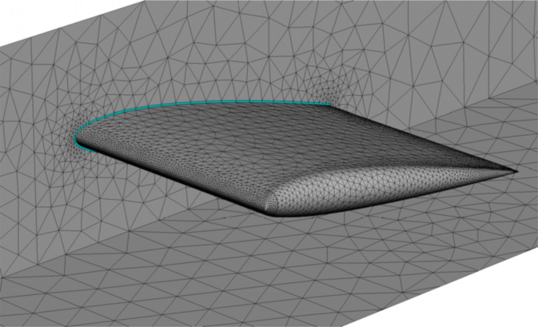}
\includegraphics[height=0.3\columnwidth]{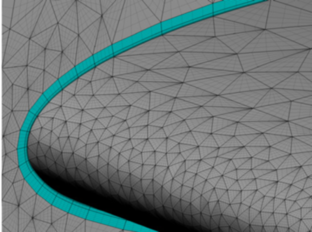}
    \caption{Surface mesh of a NACA 0012 wing geometry at polynomial order $P$= 5. The coarse prismatic boundary layer is highlighted in blue (left) and the refined boundary layer which uses three layers and a progression ratio $r$= 3 (right) (from the reference \cite{MOXEY2015636})}
    \label{naca0012mesh}
\end{figure}
 For complex  three-dimensional geometries, the generation of high-order curved 
 meshes along solid surfaces is a challenging topic 
 \cite{MOXEY2015636,NME:NME397,TURNER2016340,MOXEY2016130,FLD:FLD3767}. 
 Approaches for generating unstructured high-order meshes, extending
coarse linear grids to incorporate high-order curvature, are a relatively new 
development\cite{NME:NME397,dey1999,NME:NME2579}. The main challenge is 
robustness, since near-wall curvature must be introduced in such a way as to 
prevent the generation of self-intersecting elements \cite{MOXEY2015636}. 
Recently, several 
methods have been proposed to cope with high-order curvilinear boundary layer 
meshing, namely (1) the isoparametric approach \cite{MOXEY2015636}; (2) the 
variational approach \cite{TURNER2016340}, and; (3) the thermo-elastic analogy 
\cite{MOXEY2016130}. 
In the isoparametric approach, a reference element is mapped to a physical 
element with the supplied curvature information. The generation of a boundary 
layer mesh is achieved by splitting the reference element along the wall normal 
coordinate and then applying the mapping from a reference element to a physical 
element yielding the desired boundary layer mesh as illustrated in 
Fig. \ref{homesh}. In Fig. \ref{naca0012mesh}, a mesh for the NACA0012 aerofoil profile is shown 
which is generated by the  isoparametric approach (see \cite{MOXEY2015636} for 
details). In the variational approach, a functional which defines a measure of 
energy over mesh and takes the mesh displacement and its derivatives as its 
arguments, is minimised using a nonlinear optimisation strategy 
\cite{TURNER2016340}. It is demonstrated that the variational framework is 
efficient for both mesh quality optimisation and untangling of invalid meshes. 
In order to circumvent the drawback of elastic-model based methods which are 
used to tackle element self-intersection, a thermo-elastic analogy, as an 
extension of the elastic formulation, is proposed to `heat' or `cool' elements. 
The thermo-elastic formulation leads to an additional degree of robustness, 
which shows the potential to significantly improve  meshing \cite{MOXEY2016130}.  
{\it NekMesh},  which is a mesh generation and manipulation utility bundled 
with {\it Nektar}++, supports these strategies for high-order mesh generation.

%% file: sec4.tex
\section{Applications}
In this section, we illustrate some typical applications of spectral/{\it hp} element methods in fluid mechanics, typically in hydrodynamics. Some examples will be given through use of the pre-written solvers in {\it Nektar}++.
\subsection{The incompressible Navier-Stokes equations}
We first introduce the incompressible Navier-Stokes equations on a bounded 
domain $\Omega$ which allows one to solve the governing equations for viscous 
Newtonian fluids governed by
\begin{subequations}\label{ns}
 \begin{align}
\frac{\partial \mathbf{u}}{\partial t}+\mathbf{u}\cdot\nabla\mathbf{u}&=-\nabla p+\nu\nabla^2\mathbf{u}, \\
\nabla\cdot\mathbf{u}&=0,
 \end{align}
\end{subequations}
with appropriate boundary conditions
In {\it Nektar}++, a velocity correction scheme is employed, which uses a 
splitting/projection method where the velocity and the pressure are decoupled 
\cite{KARNIADAKIS1991414}. In the original approach, a stiffly stable time integration scheme was proposed. Briefly, high-order splitting schemes comprise of three steps involving explicit eval-uation of the non-linear terms, followed by the implicit solution of the pressure Poisson system and finally solving a series of Helmholtz problems to enforce the viscous terms and velocity boundary conditions. Time integration is handled using a generic time-stepping framework \cite{VOS2011}, utilising one of a number of implicit-explicit (IMEX) time-integration schemes.

\subsection{Transitional flows}
 In the past few decades, the spectral/{\it hp} element method has been applied to 
 laminar and transitional flows ({\it e.g.} 
 \cite{KARNIADAKIS1991414,FISCHER199469,SHERWIN1995189,SHERWIN199614,FISCHER199784,tomboulides_orszag_2000}).
  As discussed earlier, the advantages of the spectral/{\it hp} element method 
  are its low numerical dissipation and dispersion errors, which allow 
 individual flow structures to be accurately captured over long time periods. 
 Combined with 
 the development of effective stabilisation techniques  ({\it e.g.}, SVV or 
 filter-based stabilisation \cite{FISCHER2001265}), the spectral/{\it hp} 
 element method has in most recent years proved effective in modelling 
 turbulent flow 
 \cite{kirby2002,WASBERG20097333,BLACKBURN2003610,Iliescu2003,Ohlsson2011,beck2014}.

Transitional problems are those in which turbulence dominates the flow domain, 
or in which the transition to turbulence is a fundamental aspect of the 
simulation.  Here we provide a survey of the range of studies undertaken using 
the spectral/hp element method in relation to transitional flows, focusing on
(1) turbulence; (2) separated flows; (3) hydrodynamic stability, and; (4) 
vortical flows and high Reynolds number wingtip vortices. 
\begin{figure}[h!]
  \centering
\includegraphics[height=0.23\columnwidth]{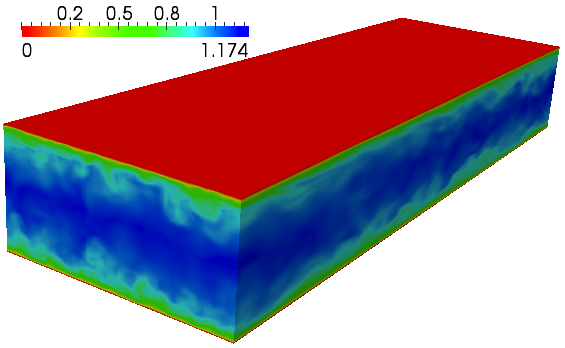}\hskip1cm
\includegraphics[height=0.23\columnwidth]{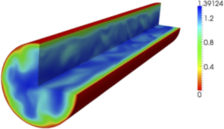}
\caption{Turbulent channel flow (left) and  turbulent pipe flow (right)}.
  \label{channelpipe}
\end{figure}

\subsubsection{Turbulent flows}
As noted in \cite{CANTWELL2015205}, highly resolved turbulent simulations were 
traditionally undertaken using spectral methods, imposing strong geometrical 
restrictions. Spectral/{\it hp} element methods alleviate this although, as 
discussed in \S 3, when the domain of interest has a geometrically homogeneous 
direction, a combination of the spectral/{\it hp} element method and the
traditional spectral method is still particularly advantageous 
\cite{BOLIS201617,BLACKBURN2004759}.

A benchmark simulation of turbulent flow over a periodic hill has been 
performed using {\it Nektar}++. It is challenging to resolve accurately due to 
the detachment of the fluid from the smooth surface and the generation of a 
recirculation region downstream of the hill \cite{CANTWELL2015205}. A 
two-dimensional mesh of 3626 
quadrilateral elements with $P=6$ in the streamwise direction  was constructed, 
and a Fourier pseudo-spectral method with 160 collocation points in the 
spanwise direction was employed to perform the simulation at a Reynolds 
number 2800.  Excellent agreement with the benchmark statistics was obtained.  
With the quasi-3D solver, turbulent channel and pipe flows have also been 
accurately simulated and validated \cite{BOLIS201617}; these are illustrated in 
Fig. \ref{channelpipe}  and are used as benchmark cases in 
{\it Nektar}++. 

\begin{figure}[h!]
  \centering
\includegraphics[height=0.17\columnwidth]{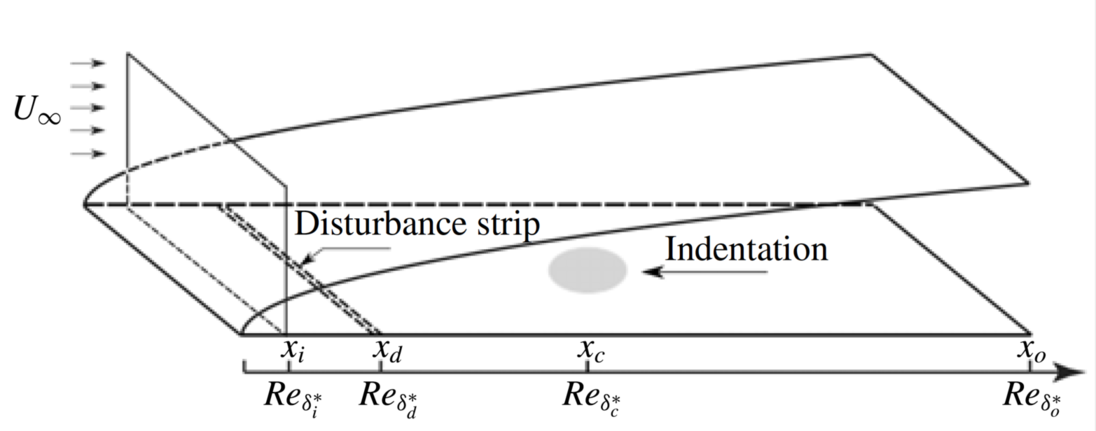}\hskip0.5cm
\includegraphics[height=0.17\columnwidth]{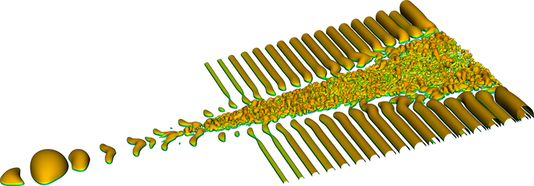}
\caption{An illustration of the computational domain for the turbulent simulation in a boundary layer (left) and  laminar-turbulent transition triggered by TS instability in a boundary layer. The iso-surfaces indicate the different pressure levels (from the reference \cite{xu_mughal_gowree_atkin_sherwin_2017})}.
  \label{boundarylayer}
\end{figure}
Recently, a global mapping technique has been developed which allows the 
simulation of fluid flows with geometrically periodic variation along 
homogeneous directions \cite{SERSON2016243,xu_mughal_gowree_atkin_sherwin_2017}. A 
typical example is to simulate turbulent flows induced by a localised surface 
deformation in the boundary layer \cite{xu_mughal_gowree_atkin_sherwin_2017}. The 
computational domain is illustrated in Fig. \ref{boundarylayer}(left) and the 
turbulent flow which forms dowstream of the surface deformation is shown in Fig. 
\ref{boundarylayer}(right). The freestream unit Reynolds number is $1.2\times10^6$ 
and the reference free-stream velocity is 18 m/s. The laminar-turbulent transition 
is initialised by the nonlinear process of upstream Tollmien-Schlichting (TS) 
disturbance breakdown to turbulence downstream of the surface indentation. The 
disturbance is generated by a disturbance strip as illustrated in Fig. 
\ref{boundarylayer}(left), which has a vibration frequency of 172Hz and is used to mimic 
the TS disturbance by the receptivity mechanism. The numerical calculations of 
this case are validated with an experimental study for which excellent 
agreement is obtained.


\subsubsection{Separated flows}

Flow separation can be one of the most important topics in fluid mechanics, due 
to its relevance to aerodynamic performance in many engineering applications. 
There are typically two kinds of separations -- geometrical, where the flow 
separates from a sharp obstacle in the flow; and pressure-induced, depending on 
the pressure gradient over a smooth surface. For both kinds of 
separation, the spectral/{\it hp} element method has been used to accurately 
capture detachment and reattachment points, {\it e.g.} 
laminar flow in a channel expansion \cite{PATERA1984468}, turbulent separation 
in a three-dimensional diffuser \cite{malm_schlatter_henningson_2012}, the flow 
around a wall-mounted square cylinder \cite{vinuesa2015}, flows over periodic 
hills \cite{diosady2014}, small separation bubbles induced by localised 
imperfections \cite{xu_mughal_gowree_atkin_sherwin_2017,xu2016}. Practically, 
the studies of separated flows are always accompanied by turbulent flows due to 
the occurrence of sensitive instabilities induced by the geometrical 
discontinuity.

Recently, with the global mapping technique, direct numerical simulations of 
the flow around wings with spanwise waviness were investigated to  explore its 
effect on the wing performance 
\cite{SERSON2017117,serson_meneghini_sherwin_2017}. The geometries are based on 
a NACA0012 profile with a small modification to obtain zero-thickness trailing 
edge to overcome meshing challenges. The wavy geometries were obtained by 
applying the following coordinate transformation to the straight infinite wing
\begin{equation}
\bar{x}=x-\frac{h}{2}\cos\left(\frac{2\pi}{\lambda}z\right),
\end{equation}
\begin{figure}[h!]
  \centering
\includegraphics[height=0.2\columnwidth]{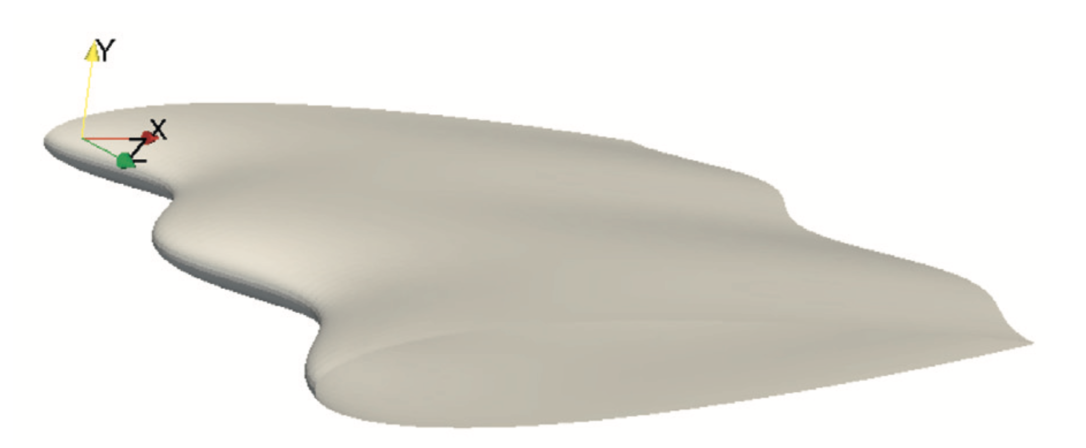}
\includegraphics[height=0.3\columnwidth]{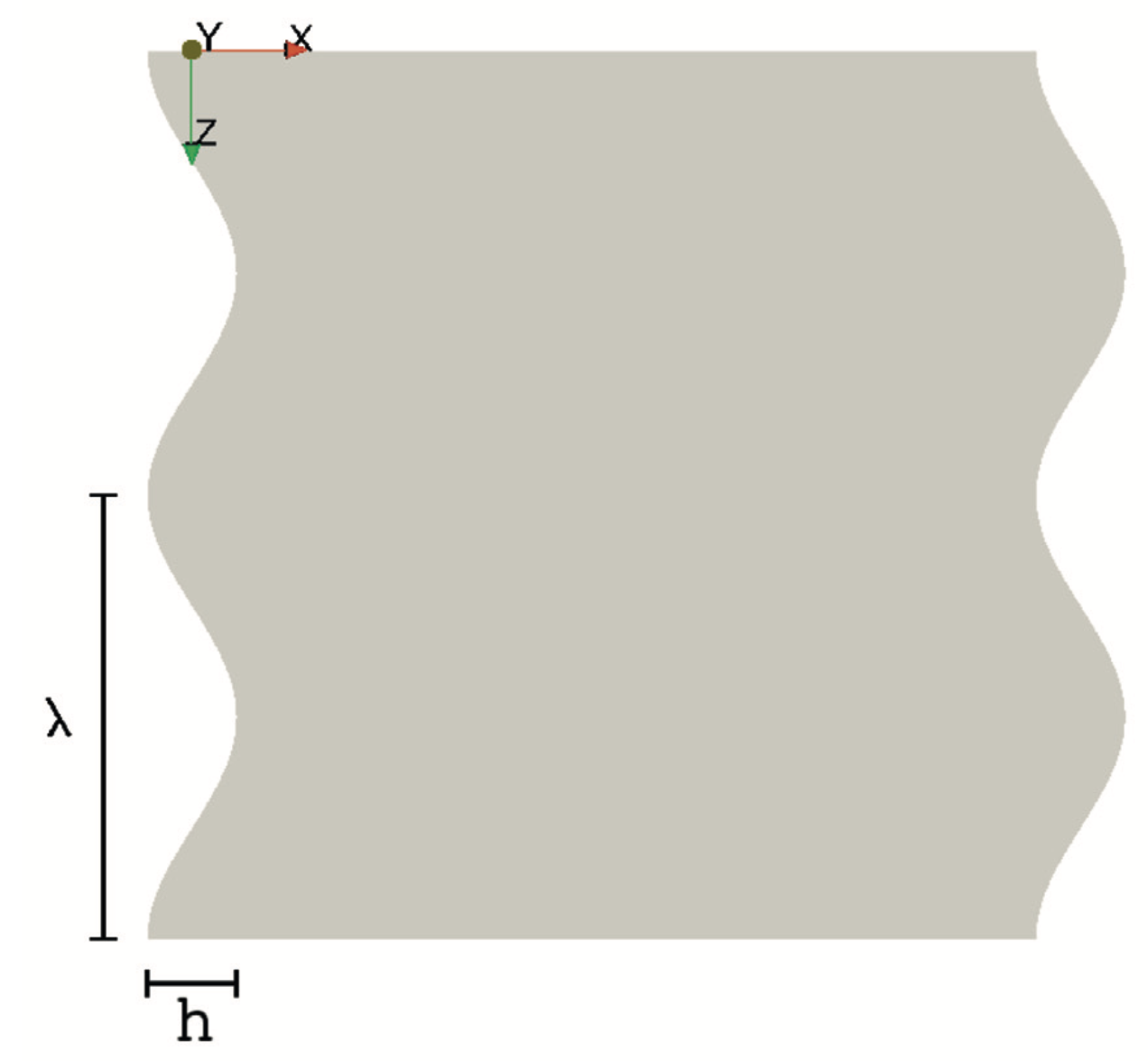}\\
(a) Perspective view \hskip 2cm (b) Planform
\caption{The geometry of a wavy wing with $h/c$=0.1 and $\lambda/d$=0.5 (from the reference \cite{SERSON2017117})}.
  \label{wings}
\end{figure}
\begin{figure}[h!]
  \centering
\includegraphics[height=0.16\columnwidth]{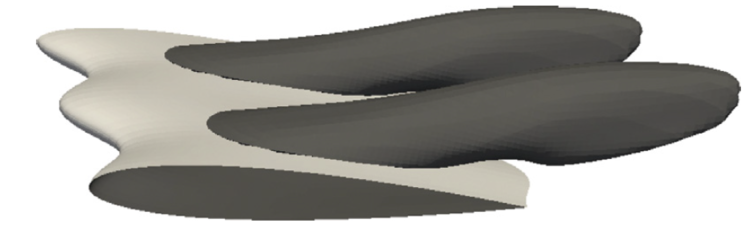}
\includegraphics[height=0.2\columnwidth]{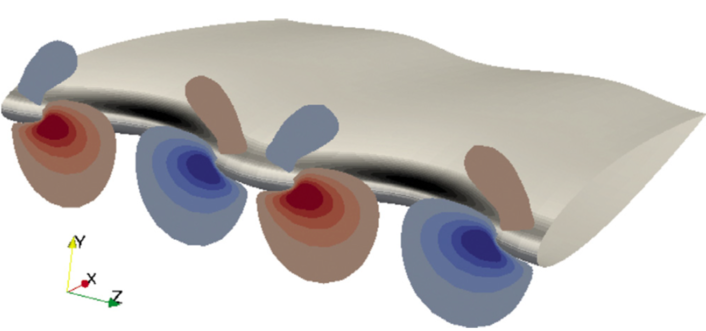}\\
(a) Recirculation zones \hskip 2cm (b) Contours of spanwise velocity
\caption{Flow properties around the wing with an attack angle $12^\circ$, $h/c$=0.1 and $\lambda/d$=0.5 (from the reference \cite{SERSON2017117})}.
  \label{wings2}
\end{figure}
where $h$ is the waviness peak-to-peak amplitude, $\lambda$ is its wavelength, $\bar{x}$ is the physical coordinate in the cord direction, and $x$ and $z$ are the chordwise and spanwise coordinates in the computational domain in Fig. \ref{wings}. At a very low Reynolds number 1000, for moderate angles of attack, the waviness leads to decreases in both drag and lift forces, which lead to a decrease in the lift- to-drag ratio and a suppression of the fluctuating lift coefficient \cite{SERSON2017117}. The physical mechanism behind these were explained. In  Fig. \ref{wings2}(a), the recirculation regions are shown and the visualisation was made considering the regions where streamwise velocity is negative. In Fig. \ref{wings2}(b), the coloured contours represent spanwise velocity at the plane $x$=-0.03, which is close to the location of the leading edge in the peak of the waviness at $x$=-0.05, depicting how the flow moves away from the waviness peak in the lower portion of the wing and it moves towards  it in the upper part.
A further study was undertaken at Reynolds numbers 10,000 and 50,000 for different attack angles through highly resolved direct numerical
simulations \cite{serson_meneghini_sherwin_2017}, which provides a better understanding of wing performance with the use of spanwise waviness.

\subsubsection{Hydrodynamic stability}

In addition to solving the fully non-linear incompressible Naiver-Stokes equations in {\it Nektar}++, the solution of the linearised incompressible Navier-Stokes equations is also supported to enable global flow stability analyses to be performed with respect to a steady or a time-periodic base flow. This process identifies whether such flows are susceptible to a fundamental change of state when an infinitesimal disturbance is introduced. The linearised incompressible Navier-Stokes equations take the form
\begin{subequations}\label{lns}
\begin{align}
  \frac{\partial \mathbf{u'}}{\partial t} + \mathbf{U} \cdot  \nabla \mathbf{u'}+\mathbf{u'} \cdot \nabla \mathbf{U} &= -\nabla p + \nu \nabla^2 \mathbf{u'} + \mathbf{f}, \\
 \nabla \cdot \mathbf{u'}&=0,
\end{align}
\end{subequations}
where $\mathbf{u}^\prime$ is the perturbation and
$\mathbf{U}$ denotes the base flow or the time-dependent periodic flow sampled at regular intervals and interpolated. For stability analysis, suitable 
boundary conditions should be imposed. The linear evolution of a perturbation 
under Eq. (\ref{lns}) can be expressed as
\begin{equation}
\frac{\partial \mathbf{u}^\prime(t)}{\partial t}=\mathscr{A}(t)\mathbf{u}^\prime(t),
\end{equation}
with an initial condition $\mathbf{u}^\prime(0)$. If the base flow $\mathbf{U}$ is steady, the perturbation $\mathbf{u}^\prime(t)$ be expressed by eigenmode solutions of $\mathscr{A}(t)$  as follows
\begin{equation}
\mathbf{u}^\prime(\vec{x},t)=\exp(\lambda_j t)\tilde{\mathbf{u}}_j+{\rm c}.{\rm c}.
\end{equation}
The dominant eigenvalues and eigenmodes of the operator $\mathscr{A}(t)$ are defined by solutions to the following equation
\begin{equation}
\mathscr{A}(t)\tilde{\mathbf{u}}_j=\lambda_j\tilde{\mathbf{u}}_j,
\end{equation}
The leading eigenvalue $\lambda_j$  is used to detect the global stability of 
the flow. Using a similar approach, the operator $\mathscr{A}^*(t)$ for the 
adjoint form of the linearised Navier-Stokes evolution operator can be used to examine the receptivity of the flow and, in combination with the 
direct mode, identify the sensitivity to base flow modification and local 
feedback. The {\it direct} and {\it adjoint} methods can be combined to explore 
convective instabilities over different time horizons $\tau$  in a flow by 
computing the leading eigenmodes of $(\mathscr{A}^*\mathscr{A})(\tau)$ known as transient growth analysis.
The eigenvalues and eigenmodes of $\mathscr{A}(\tau)$ and  
$(\mathscr{A}^*\mathscr{A})(\tau)$ can be solved directly if the dimensions of 
the matrices are small. Practically, time-stepper-based methodologies based on Arnoldi iteration methods  (which are now supported 
in {\it Nektar++}) \cite{barkley2008}, can be more efficient for complex 
problems \cite{BLACKBURN2008,BLACKBURN2008B,CANTWELL2010A,CANTWELL2010B} than 
direct calculation of the eigenvalue spectrum from the linearised Navier-Stokes 
operator.

\begin{figure}[h!]
  \centering
\includegraphics[width=0.45\columnwidth]{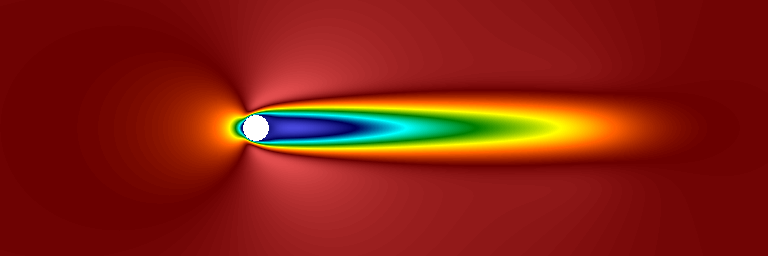}
\includegraphics[width=0.45\columnwidth]{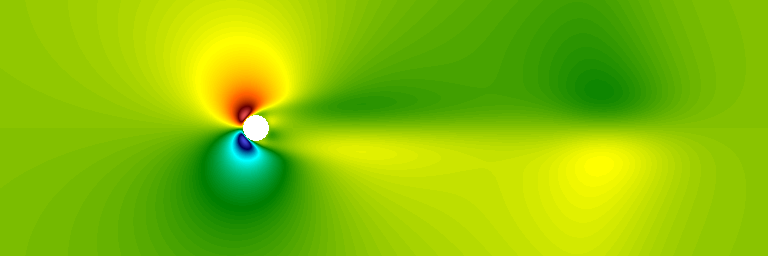}\\
(a) $u$ \hskip 5cm (b) $v$\\
\includegraphics[width=0.45\columnwidth]{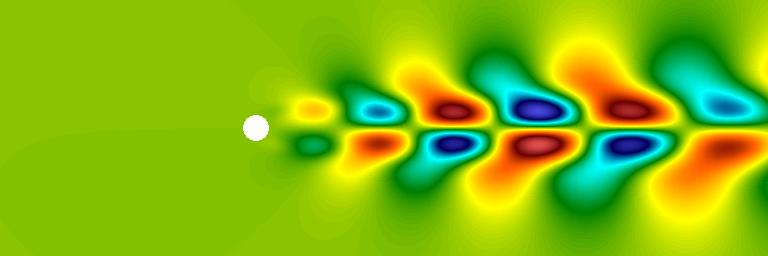}
\includegraphics[width=0.45\columnwidth]{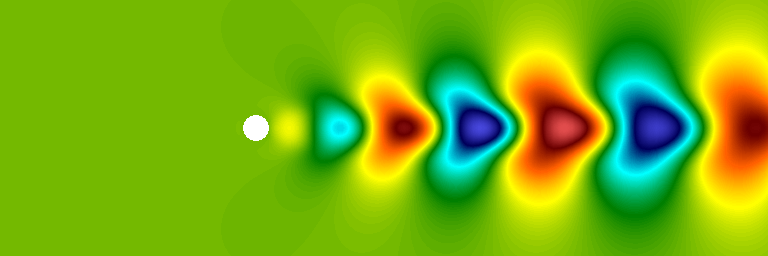}\\
(c) $\tilde{u}$ \hskip 5cm (d) $\tilde{v}$
\caption{Direct stability analysis of two-dimensional flow past a circular cylinder
at $Re$=50: (a) the x-component of the base flow $\mathbf{u}$; (b) the y-component of the base flow $\mathbf{u}$; (c) the x-component of the dominant direct mode;(d) the y-component of the dominant direct mode. }.
  \label{cylinder}
\end{figure}
To illustrate the direct stability analysis with a classic example, we 
calculated the dominant mode of 
the two-dimensional flow past a circular cylinder at $Re=50$. We show the two 
components of the base flow in Figs. \ref{cylinder} 
(a) and (b) and  the two components of the dominant direct mode in Figs. 
\ref{cylinder} (c) and (d). The leading eigenmode is characterised by the 
asymmetry in 
the streamwise component and symmetry in the cross-stream component. We also 
observe the spatial distribution of the modes, with the leading direct modes 
extending far downstream of the cylinder.
Recently, the instability methods using the spectral/{\it hp} element method have been employed for investigating the stability analysis of vortical flows and controlling wakes of flows past bluff bodies \cite{Broadhurst2006,rocco2014,jordi2015}. With the linearised Navier-Stokes equations, the interaction between instability waves and surface distortion in a 2-D and 3-D boundary layer have been precisely investigated \cite{xu2016,xu_lombard_sherwin_2017,xu_mughal_gowree_atkin_sherwin_2017}.

\subsubsection{Vortical flows and wingtip vortex}

Simulating and understanding vortical flows are vital in hydrodynamics and 
aerodynamics. In many applications,  manipulation of vortices in the vicinity 
of flow boundaries is crucial for improving performance in engineering practice 
\cite{lombard2016,devenport_rife_liapis_follin_1996,arndt2002,ford_babinsky_2013,leveke2016,feys_maslowe_2016}.
Developing a better understanding of the near wake of the vortex, lying within 
one chord length of the trailing edge, is essential in understanding the 
complex flow structure interactions \cite{lombard2016}, which is also crucial 
in understanding cavitation in vortical structures generated by hydro 
propellers \cite{arndt2002}. Computationally, in high Reynolds number flows, 
accurate numerical simulations of these kinds of vortical structures are 
challenging for the traditional numerical methods due to numerical 
dissipation.  A recent study in which wingtip vortices were simulated 
demonstrated that the 
adjustable and controllable low-dissipation properties of the spectral/hp 
element method were beneficial for modelling and 
simulating vortical flows \cite{lombard2016,MOURA2015695,MOURA2017615}. 

To illustrate the ability of the spectral/{\it hp} element method to accurately 
capture vortical flows at a Reynolds number $1.2\times 10^6$, a rectangular 
wing with a NACA0012 profile is investigated with a rounded wing cap 
(consequently, a longer semispan where the wing is thickest) and a blunt 
trailing edge \cite{lombard2016}. The results showed better correlation with 
experimental results than previous numerical results, both in terms of the 
static pressure distribution, prediction of the jetting velocity, vortex 
spanwise location, and the ability to resolve the secondary vortex interaction 
with the main wingtip vortex. The iLES 
method based on SVV  has been shown to be a compelling alternative for 
computing complex  vortex-dominated flows, such as the wingtip vortex, 
motivating its use for complex industrially relevant cases where high-fidelity 
computational fluid dynamics can become an enabling technology.

\begin{figure}[h!]
  \centering
\includegraphics[width=0.6\columnwidth]{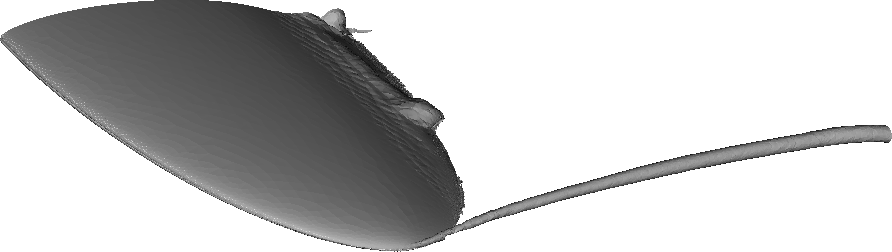}
\caption{Tip vortex of an elliptic hydrofoil}.
  \label{tipvortex}
\end{figure}

\begin{figure}[h!]
  \centering
\includegraphics[width=0.3\columnwidth]{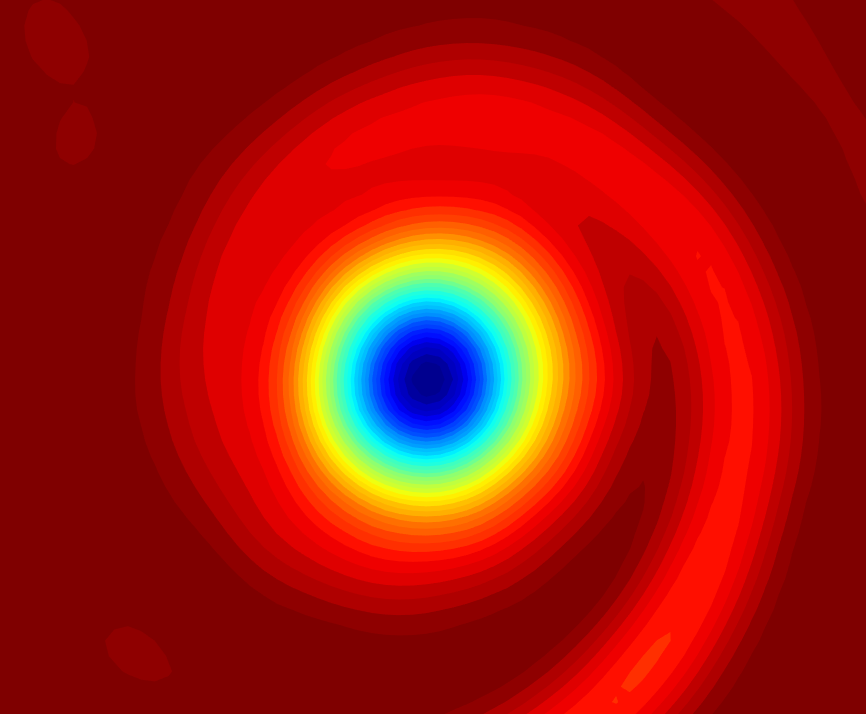}
\includegraphics[width=0.3\columnwidth]{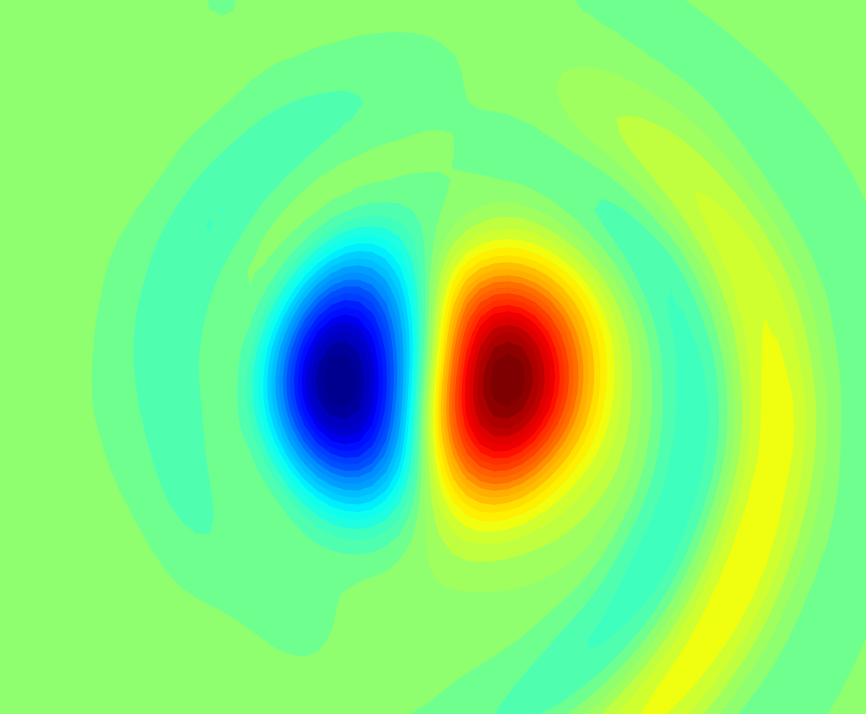}
\includegraphics[width=0.3\columnwidth]{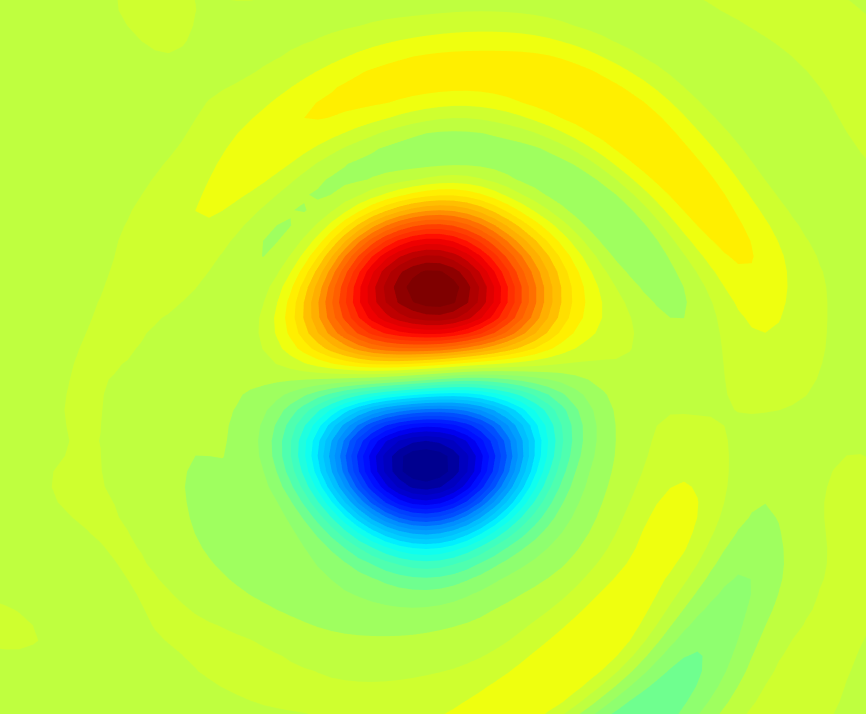}
\caption{Vorticity components $\omega_x$, $\omega_y$ and $\omega_z$ along axial (left), tangential (mid) and spanwise (right) directions at the section $x/c=1.06$}.
  \label{tipvort}
\end{figure}
Built on the confidence of simulating the NACA0012 wingtip vortex, recently, an initial simulation with a Reynolds number $1.217\times10^6$ is currently being conducted to simulate the tip vortex of an elliptic hydrofoil where the incident flow attack angle is $7^\circ$. The tip vortex structure is illustrated by the iso-surfaces of the total pressure in Fig. \ref{tipvortex}. The contour figures of the vorticity components are shown at the section $x/c=1.06$ in Fig. \ref{tipvort} and the origin is located at the tip of the hydrofoil. In Fig. \ref{tipvort} (left),  the vortex core is shown in axial vorticity $\omega_x$  and tangential and spanwise vorticity components ($\omega_y$ and $\omega_z$) are shown in dipoles in Fig. \ref{tipvort} (mid) and (right).   This calculation will be used to provide vortex core pressure for studying cavitation in a vortex structure.  The numerical results will be going to be compared with experimental data.

\subsection{Waves in ocean engineering}

This section is devoted to outlining the recent progress in spectral/{\it hp} element modelling of wave propagation and wave-body interaction. For water waves propagation the fully nonlinear potential flow (FNPF) equations is the fundamental governing equation and can be derived from the Navier-Stokes equations by assuming inviscid and irrotational flow. One of the main challenges in the last decade have focused on use of spectral/{\it hp} element methods for coastal engineering \cite{EESHB05}, and the development of robust and efficient solvers with support for unstructured meshes to capture realistic shore lines, geometric features, and adapt meshes to relevant features of the solution \cite{Brocchini2013}.

\subsubsection{Fully nonlinear potential flow }
Finite elements are widely used for solving the FNPF equations \cite{WuEatockTaylor1994,CaiEtAl1998,MaEtAl2001,MaYan2006}, but the use of spectral/{\it hp} elements remains scarse. The first attempt to solve the FNPF equations using spectral/{\it hp} elements is due to Robertson and Sherwin \cite{RobertsonSherwin1999} using an Arbitrary Lagrangian Eulerian (ALE) approach for the free water surface in a two-dimensional setting. Solving the FNPF equations is non-trivial, due to the need to evolve a set of highly nonlinear free surface boundary conditions in a robust way, together with the efficiency solution of a Laplace problems -- possibly -- in large ocean areas. For spectral/{\it hp} element models, Robertson and Sherwin identified a mesh asymmetry problem associated with triangulation of the fluid domain, which gives rise to real-valued eigenvalues and thus represented a severe instability problem. This was handled by adding a diffusive term in the kinematic free surface boundary condition proportional to the mesh asymmetry, but at the cost of reduced convergence rates. The problem of mesh asymmetry which destroys the dispersion relation was circumvented by Engsig-Karup {\em et al.} \citep{EngsigKarupEtal2016,EngsigKarupEtal2016b} by solving the FNPF equations in a $\sigma$-transformed domain \cite{CaiEtAl1998} using a single layer of quadrilaterals in 2D and prisms in 3D. The FNPF model was shown to exhibit exponential convergence even for steep nonlinear stream function waves which most simpler wave models cannot handle. To stabilise the nonlinear simulations, it was found that the quartic nonlinear terms in the free surface equation need to be integrated exactly and for the steep waves a mild (1\%) modal filter could be used to remove high frequency noise that can stem from aliasing in nonlinear terms for marginal resolution, and the gradient recovery steps that reconstruct the gradients of the solution (e.g. velocities) from the $\mathcal{C}^0$ approximations. 

\begin{figure}
\begin{center}
\includegraphics[width=0.8\columnwidth]{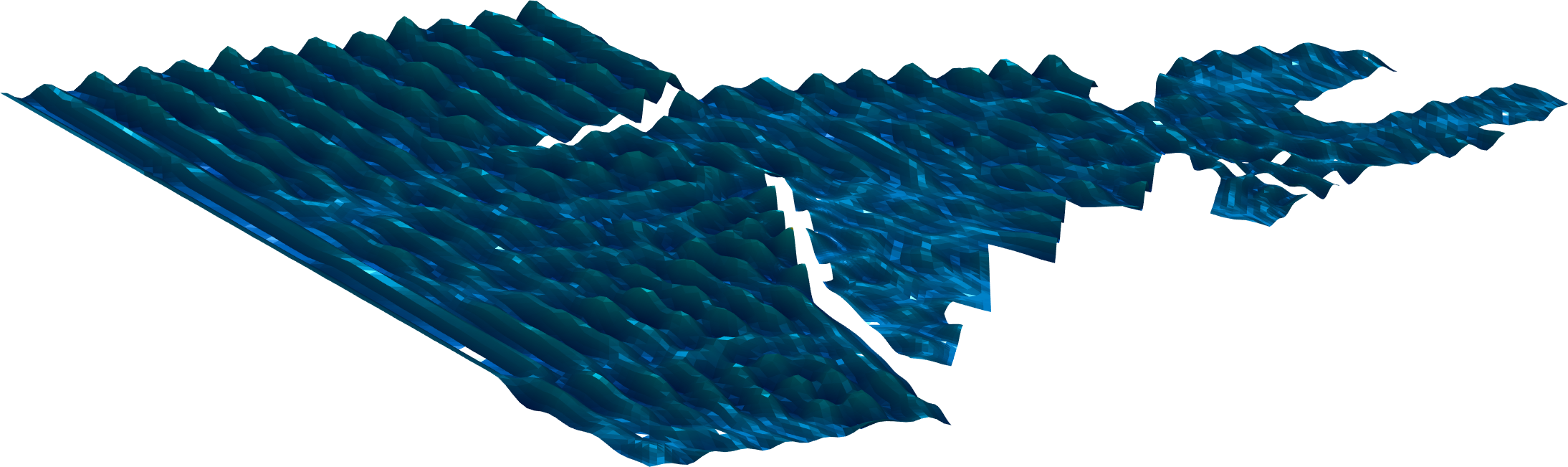}
\end{center}
\caption{Snap shot of free surface elevation in the Port of Visby using fifth order method.}
\label{fnpfVisby}
\end{figure}

\begin{figure}
\begin{minipage}{0.33\columnwidth}
\centering (a)\\
\includegraphics[width=1.1\columnwidth]{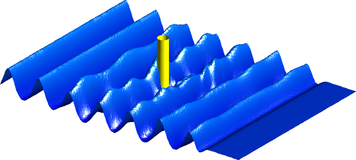}
\end{minipage}
\begin{minipage}{0.33\columnwidth}
\centering (b)\\
\includegraphics[width=1.1\columnwidth,]{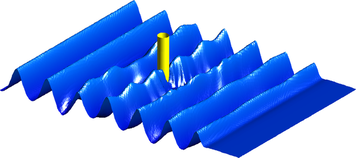}
\end{minipage}
\begin{minipage}{0.33\columnwidth}
\centering (c)\\
\includegraphics[width=1.1\columnwidth]{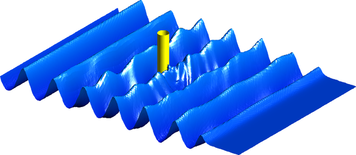}
\end{minipage}
\caption{Wave scattering of regular waves by a cylinder. Wave propagating from left to right. (a) Maximum wave run-up on front-side of cylinder, (b) wave passing cylinder and (c) wave run-up on lee-side}
\label{fnpfFerrant}
\end{figure}

Letting $\bar{\nabla}$ denote the horizontal gradient operator, the velocity potential $\phi$ satisfy the Laplace equation
\begin{equation}
\bar{\nabla} \phi + \frac{\partial^2 \phi}{\partial z^2} = 0\,, 
\end{equation}
with time-dependent free surface boundary conditions on $\Gamma_{z=\eta}$ given in the Zakahrov form \cite{Zakharov1968}
\begin{align}
\frac{\partial \eta}{\partial t} + \bar{\nabla}\cdot \bar{\nabla} \tilde{w} - \tilde{w}\left(1 +\bar{\nabla}\eta\cdot \bar{\nabla}\eta\right) = 0\,, \\
\frac{\partial \tilde{\phi}}{\partial t } + g \eta + \frac{1}{2}\left(\bar{\nabla}\tilde{\phi}\cdot \bar{\nabla}\tilde{\phi} + \tilde{w}^2 \left( 1 +\bar{\nabla} \eta \cdot  \bar{\nabla} \eta \right) \right) = 0\,, 
\end{align}
and zero normal flow on the bottom $\Gamma_{z=-h}$ and on structures $\Gamma_{b}$ 
\begin{equation}
\frac{\partial \phi}{\partial n } = 0\,.
\end{equation}
Here $\eta$ is the free surface elevation, $w$ the vertical velocity, $h$ the still water depth and $g$ the acceleration of gravity. The $\sigma$-transformed FNPF solver as presented in \cite{EngsigKarupEtal2016b} has been implemented in the {\it Nektar}++ framework in order to allow for larger-scale modelling \cite{MieritzEtal2018}. The solution is advanced explicitly in time by solving the free-surface equations. The computed $\tilde{\phi}^{n+1}$ are used as Dirichlet condition for the Laplace equation at $\Gamma_{z=\eta}$. After solving for the velocity potential, $\tilde{w}^{n+1}$ is obtained by a $C^0$ gradient recovery as $\tilde{w}^{n+1}$ is needed to advance the free surface conditions another time step. Explicit time-stepping schemes are effective for FNPF as the time-step restriction is not dependent on the horizontal mesh size, but only on the vertical resolution and water depth, see \cite{EngsigKarupEtal2016}. A $\sigma$-transformed FNPF solver is well-suited for large-scale wave propagation simulations, since re-meshing is avoided. Also, an intrinsic property of FNPF solvers is the ability to get accurate kinematics in all of the fluid domain as a part of the solutions, which is relevant for wave-induced load predictions on marine structures. The $\sigma$-transformed FNPF solver can be used for computing {\it e.g.} short wave disturbances in ports, see Fig.~\ref{fnpfVisby}, and wave scattering from bottom mounted vertical cylinders, see Fig.~\ref{fnpfFerrant}.

\begin{figure}
\begin{minipage}{0.33\columnwidth}
\begin{center}
(a)\\
\includegraphics[width=0.95\columnwidth]{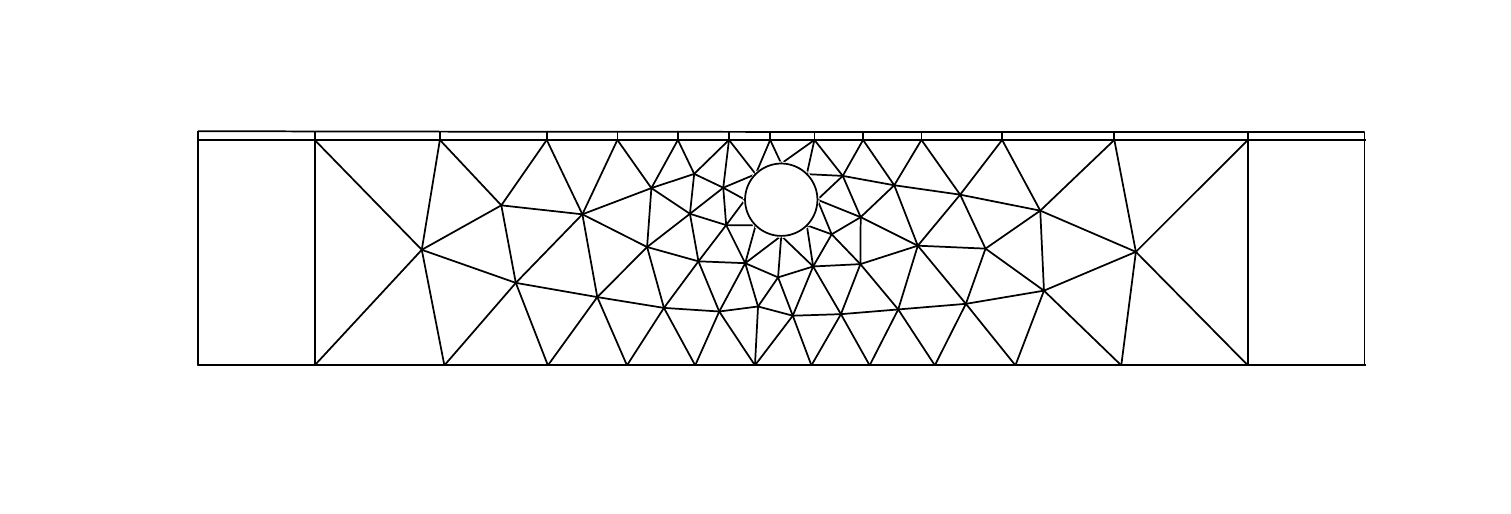}\\
\end{center}
\end{minipage}
\begin{minipage}{0.33\columnwidth}
\begin{center}
(b)\\
\includegraphics[width=0.95\columnwidth]{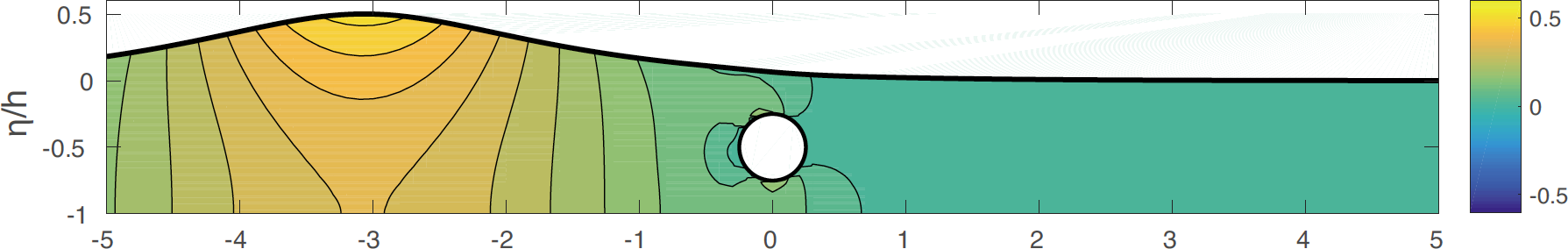}\\
\includegraphics[width=0.95\columnwidth]{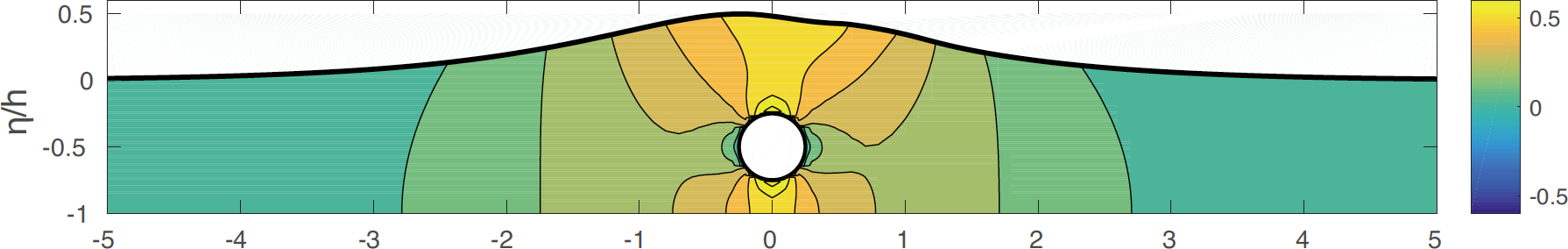}\\
\includegraphics[width=0.95\columnwidth]{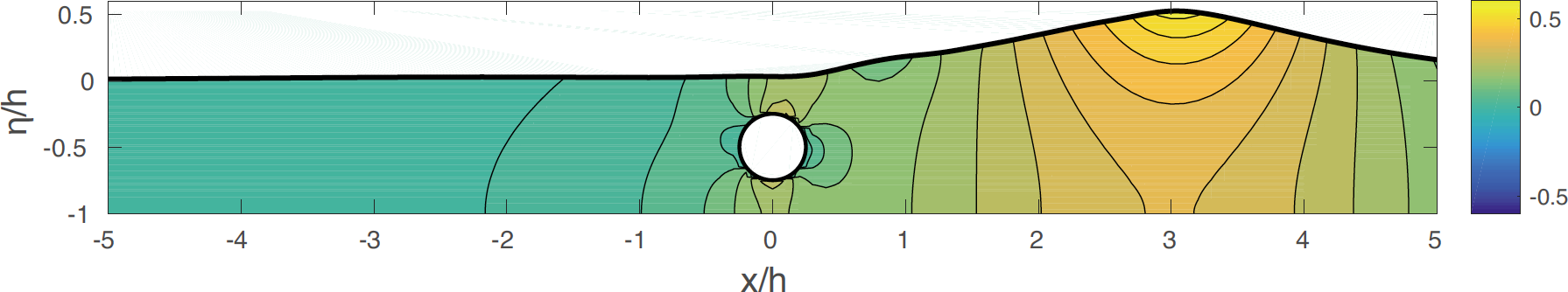}
\end{center}
\end{minipage}
\begin{minipage}{0.33\columnwidth}
\begin{center}
(c)\\
\includegraphics[width=0.95\columnwidth]{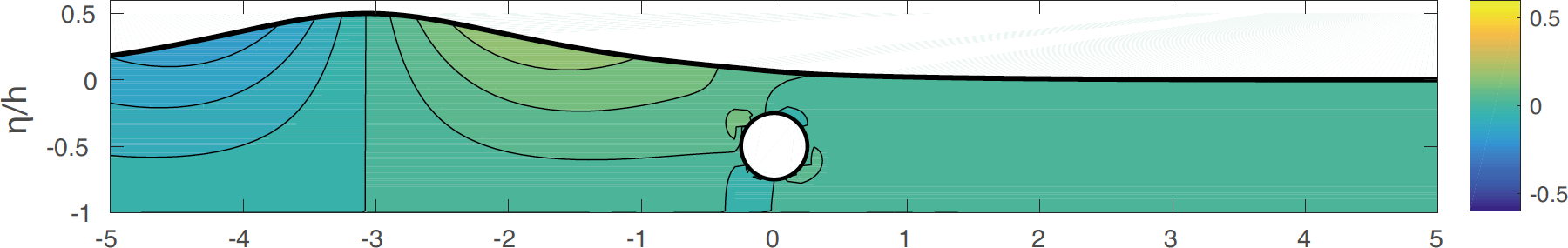}\\
\includegraphics[width=0.95\columnwidth]{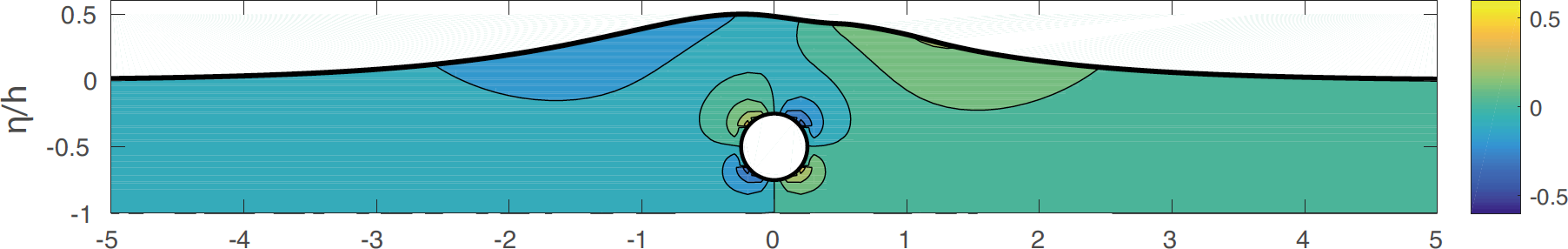}\\
\includegraphics[width=0.95\columnwidth]{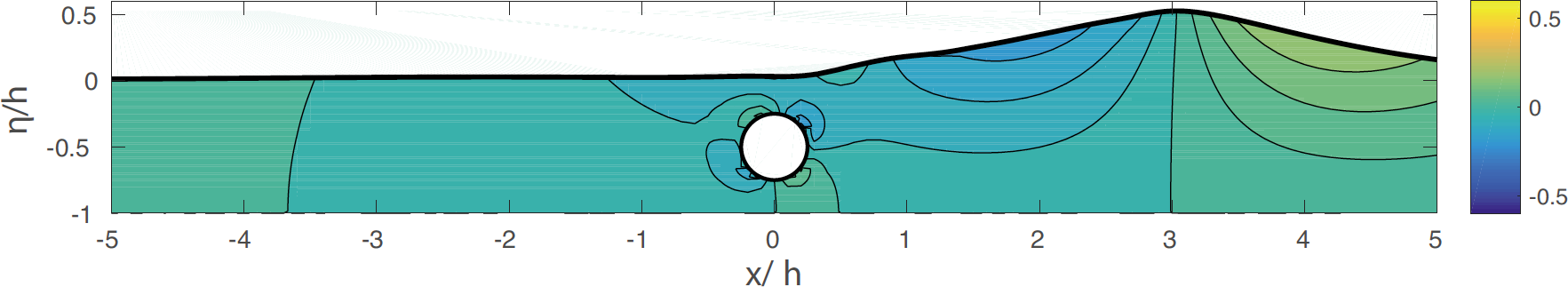}
\end{center}
\end{minipage}
\caption{Solitary wave passing a submerged horizontal cylinder. (a) zoom in on the hybrid mesh, (b) horizontal velocity and (c) vertical velocoity (from the reference \cite{APEK2017})}
\label{fnpfBody}
\end{figure}

The use of $\sigma$-transformed domains excludes any truncated bodies in the domain. In order to handle arbitrarily shaped bodies in the domain the mixed- Eulerian-Lagrangian (MEL) approach should be used and there are ongoing efforts to implement a SEM based on the MEL approach. In \cite{APEK2017} a work-around on the mesh asymmetry problem associated with the MEL was presented. By using hybrid meshes consisting of a single layer of vertically aligned quads (in 2D) at the free surface, the eigenvalues were shown to be purely imaginary for any triangulation of the inner field. When using higher-order local polynomial expansions as done in the spectral/{\it hp} element method, a local re-meshing (movement of quadrature points inside the local elements) as well as global re-meshing (movement of vertices) are needed in the MEL approach. An example of a spectral/$hp$ element simulation using MEL for wave-body interaction are illustrated in Fig.~\ref{fnpfBody}. Here a solitary wave is passing over a horizontal submerged cylinder.

\subsubsection{Shallow water wave equations}
While the application of spectral/{\it hp} element method for FNPF equations is emerging, in the field of depth-integrated shallow water long wave equations there is a wider use of spectral/{\it hp} elements methods. These equations can be derived from the FNPF equations by expanding the velocity potential in terms of the vertical coordinate and integrating the Laplace problem over the fluid depth. This removes the vertical dependence of the problem. Non-hydrostatic wave equations (such as Boussinesq-, Serre- or Green-Naghdi-type equations) are used for wave propagation and transformation in nearshore regions. The dispersive effects are included in the equations through higher-order mixed and spatial derivatives.  Eskilsson and Sherwin presented the first spectral/{\it hp} model for 1D Boussinesq-type equations \cite{ES02}, and later in 2D \cite{EskilssonSherwin2005}. In the latter study was illustrated that the use of high-order schemes is computationally efficient compared to low-order methods for long-time integration. This is general knowledge, mathematically proven in \cite{KO72}, but the importance is accenturated in wave equation where the numerical dispersion error must be kept small compared to the physical dispersion terms. Additional spectral/{\it hp} studies of weakly dispersive Boussinesq equations are \cite{EskilssonEtal2006,DumbserFacchini2016}. Engsig-Karup {\em et al}. \cite{ENG06,EHBM06,EHBW08} solved a set highly dispersive and nonlinear Boussinesq-type equations. More recently the interest has turned toward Green-Naghdi type equations due to their description of nonlinearity and a number of spectral/{\it hp} element methods have been put forward \cite{PandaEtal2014,DuranMarche2015,SamiiDawson2018}. We note that all the above models are based on high-order discontinuous Galerkin (DG) methods.

\begin{figure}
\begin{minipage}{0.49\columnwidth}
\begin{center}
(a) \\
\includegraphics[width=8cm]{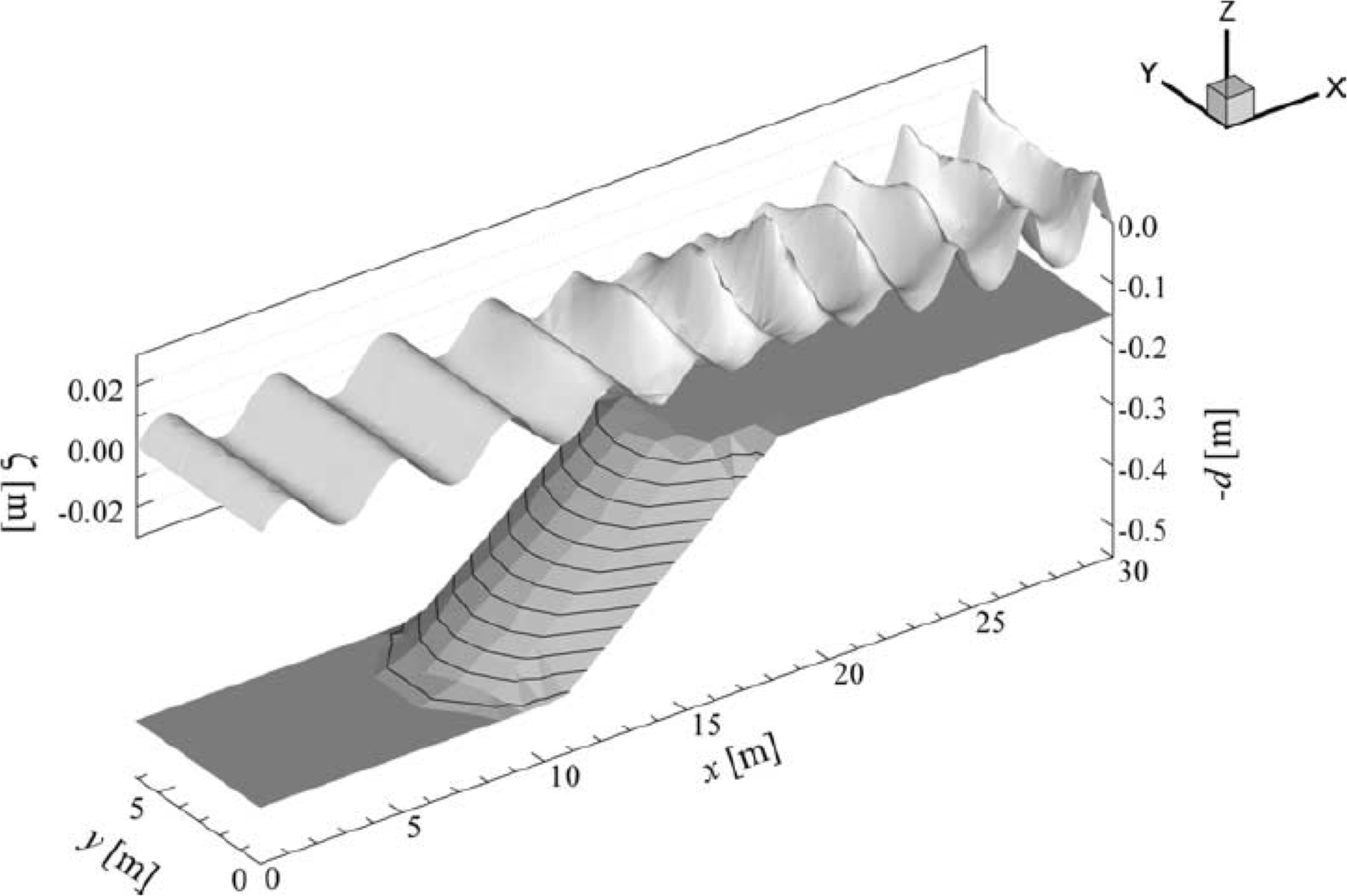}
\end{center}
\end{minipage}
\begin{minipage}{0.49\columnwidth}
\begin{center}
(b)\\
\includegraphics[width=8cm]{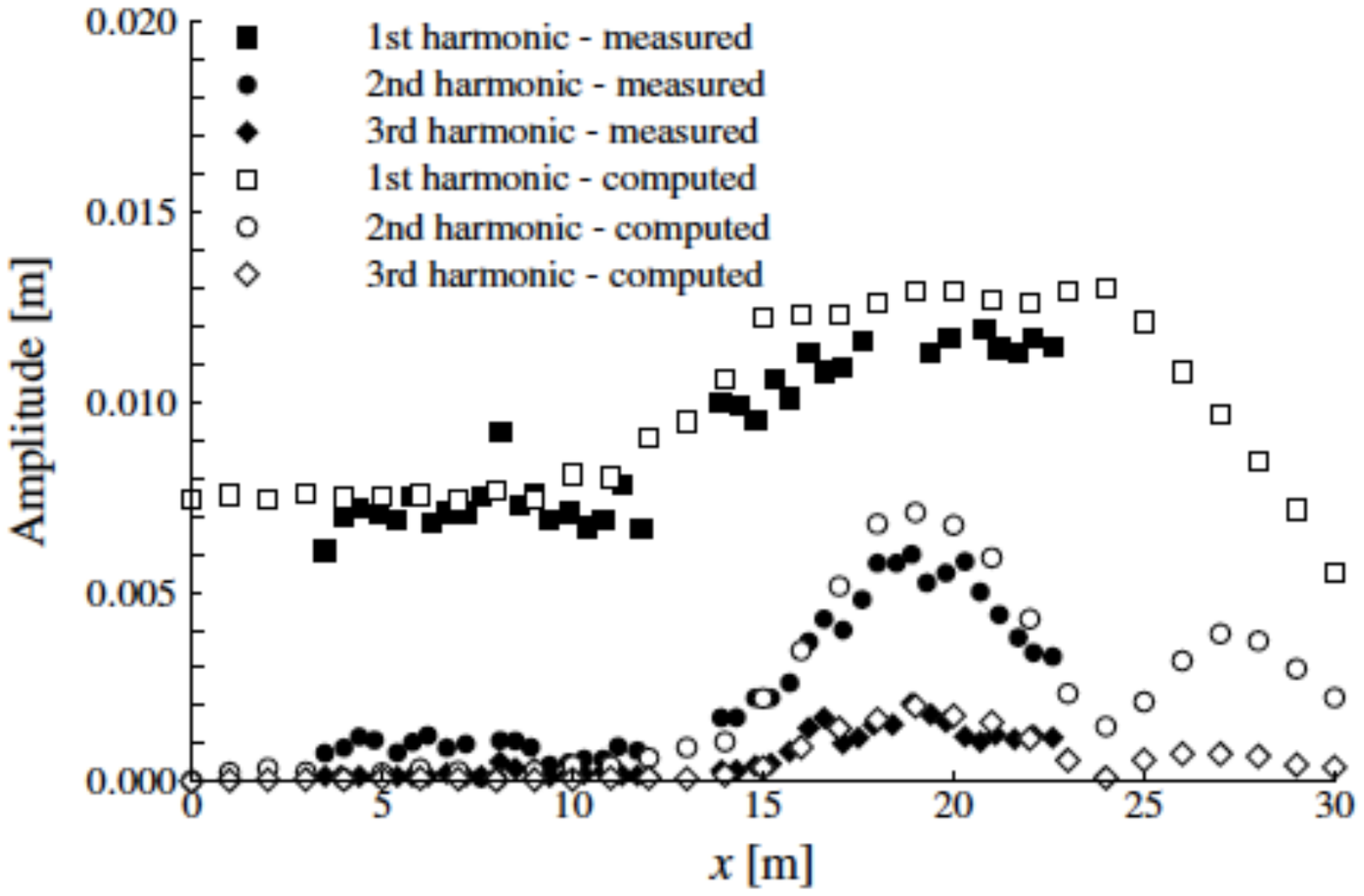}
\end{center}
\end{minipage}
\caption{Higher harmonics generation on a semi-circular shoal. (a) 3D surface view and (b) computed and measured harmonics through the centerline (from the reference \cite{EskilssonSherwin2006})}
\label{boussinesq}
\end{figure}

The classical Boussinesq equations due to Peregrine \cite{Peregrine1967} read
\begin{align}
&\frac{\partial \eta}{\partial t} + \bar{\nabla} \cdot \left(\left(h+\eta\right) \bar{\mathbf{u}} \right) = 0\,, \\
&\frac{\partial \bar{\mathbf{u}}}{\partial t} + g\bar{\nabla} \eta + \bar{\mathbf{u}} \cdot \bar{\nabla}\bar{\mathbf{u}} + \frac{h^2}{6} \bar{\nabla}\left( \bar{\nabla} \cdot \frac{\partial \mathbf{u}}{\partial t} \right) - \frac{h^2}{2} \bar{\nabla}\left(\bar{\nabla} \cdot \frac{\partial (h\mathbf{u})}{\partial t} \right) = 0\,,
\end{align}
where $\bar{\mathbf{u}}$ is the horizontal velocities. The exist a pre-written solver for the Peregrine equations in {\it Nektar}++. The equations are solved in terms of conservative variables and the solution scheme is using the wave-continuity step as introduced in \cite{EskilssonSherwin2005}. First the advection part is approximated and used in an intermediate step. In the intermediate step the dispersive terms are reformulated into  a Helmholtz equation in terms of the auxiliary variable $q=\bar{\nabla}\cdot \partial_t (d\mathbf{u})$, in which $d=h+\eta$. Subsequently, a projection back to conservative variables is performed. The Boussinesq equations can be used for computing the generation of higher harmonics. A classical example is the semi-circular shoal case, see Fig.~\ref{boussinesq}. 

With regard to non-dispersive, hydrostatic wave equations the shallow water equations (SWE) are typically used in marine applications to predict tidal forcing but can also be used for other long wave phenomena such as tsunami and storm-surge predictions. For tsunami modelling {\it hp}-adaptive DG models have been presented in order to resolve the shock wave propagation in detail \cite{BlaiseStCyr2012,BlaiseEtal2013,Bonev2017}. However, the main bulk of work on spectral/{\it hp} element methods for the SWE are linked to numerical weather prediction or ocean circulation where the SWE is used as a stepping stone towards solving the 3D primitive equations, see {\it e.g.} \cite{Ma93,IskandaraniEtal95,TaylorEtal1997,GiraldoEtal2002,NairEtAl2005,LauterEtal2008,BaoEtAl2014}. 

\begin{figure}
\begin{center}
\begin{minipage}{0.31\columnwidth}
\begin{center}
(a)\\
\includegraphics[width=0.89\columnwidth]{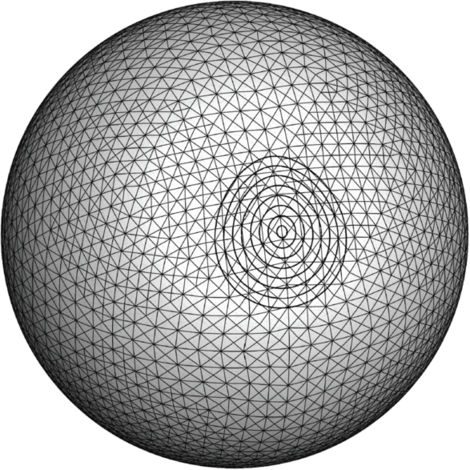}
\end{center}
\end{minipage}
\begin{minipage}{0.31\columnwidth}
\begin{center}
(b)\\
\includegraphics[width=0.89\columnwidth]{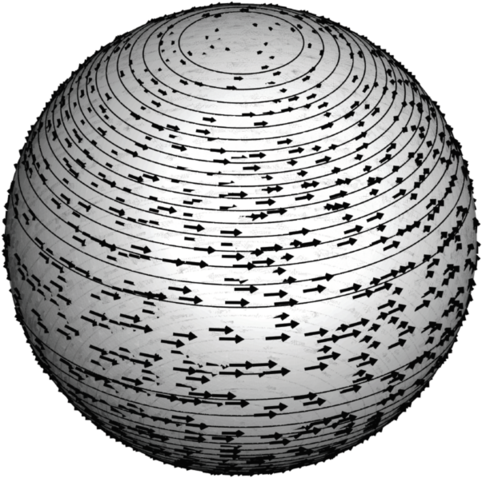}
\end{center}
\end{minipage}
\begin{minipage}{0.31\columnwidth}
\begin{center}
(c)\\
\includegraphics[width=0.89\columnwidth]{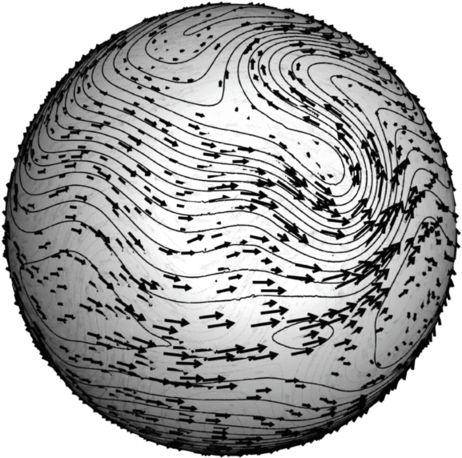}
\end{center}
\end{minipage}
\end{center}
\caption{Spherical shallow water flow over an isolated mountain. (a) mesh, (b) initial surface elevation and (c) surface elevation after 15 days. (from the reference \cite{ChunEskilsson2017})}
\label{mmfSWE}
\end{figure}

The SWE is solved in {\it Nektar}++ as a sub-model for the dispersive Boussinesq solver simply by ignoring the dispersive step. Recently a version of the SWE model capable of solving the flow on curved surfaces  using {\it Nektar}++ was presented in \cite{ChunEskilsson2017}. The model uses the so-called method of moving frames \cite{Chun2012} to express the SWE on curved and rotating geometries. The model was shown to exhibit exponential convergence and handled the standard test cases for spherical SWE, see Fig.~\ref{mmfSWE} for the flow around an isolated mountain. 

%% file: sec5.tex
\section{Future directions and perspectives}

Significant progress has been made in theoretical and applied aspects of the 
spectral/{\it hp} element method over the past two decades which have enabled 
its application to a wide range of challenging hydrodynamics and broader 
industrial problems. However, there still remain a number of challenges 
\cite{FLD:FLD3767}. In 
order to enable the spectral/{\it hp} element method to be used to address 
difficult, large-scale, high Reynolds number flow problems in complex 
geometries, we see a number of areas which still require improvement:
\begin{enumerate}
\item High-order mesh generation: Robustly generating high-order meshes with CAD
  conforming representation of the underlying geometry remains one of the key
  challenges to industrial application.
\item One of the major benefits of the spectral/{\it hp} element method is the 
opportunities presented for optimising the computational discretisation in 
accordance with the physics being modelled through {\it hp} adaptivity. {\it 
p}-adaptivity and {\it r}-adaptivity have recently been explored 
\cite{MOXEY2017TOWARDS, EKELSCHOT2017} but further effort is needed in 
identifying the optimal refinement algorithms to choose to optimise the 
computational cost while also improving the numerical accuracy of solutions.
\item The stabilisation techniques discussed in \S 2, such as SVV and
  projection-based filtering techniques, require further improvement through
  further calibration for marginal/under-resolved computations. As mentioned by Sagaut  \cite{sagaut}, ``{\it some specific numerical stabilisation procedures can be defined, which tune the numerical dissipation in such a way that the results remain sensitive to subgrid modelling"}. In terms of the similarity between artificial dissipation and the direct energy cascade model, some numerical stabilisation techniques can be used to perform ``no-model'' large eddy simulations (iLES). However, Sagaut stated that ``{\it certain studies have shown that, for coarse grids, i.e. high values of the numerical cutoff length, increasing the order of accuracy of the upwind scheme can lead to a degradation of the results} \cite{TAFTI1996647}". The proposed `DGKernel', which emulates the upwind properties of a DG scheme, still requires more detailed calibration for a range of high- Reynolds fluid flows.
\item Verification and Validation of spectral/{\it hp} element simulations for 
under-resolved computations: The
  use of stabilisation techniques, such as SVV, allows us to interpret
  the under-resolved numerical calculations as implicit large eddy
  simulation. However, we need to continue to investigate how the stabilisation methods and numerical properties interact to ensure the benefits of high-precision are not destroyed through artificial/numerical pollution.
\end{enumerate}

In the context of hydrodynamics applications, there is substantial opportunity 
for gaining an improved understanding of a wide range of physical mechanisms 
through the use of the spectral/{\it hp} element method. In particular:
\begin{enumerate}
\item Tip vortex modelling: As indicated in \cite{lombard2016},
  without the accurate modelling of the three dimensional boundary
  layer, the developing vortex remained challenging to compute
  accurately, even for the advanced RANS models correcting for the
  high degree of curvature in the flow.  In addition, accurately capturing the
  low-pressure region within the vortex core and sustaining this low
  pressure even just one chord length downstream of the trailing
  edge is particularly challenging. As demonstrated
  \cite{lombard2016}, the SVV-based iLES method has been shown to be potential alternative for computing complex unsteady vortex
  dominated flows. Simulating vortical flows using the spectral/{\it hp}
  element method with the concept of the SVV-based iLES can be
  regarded as an interesting alternative method in hydrodynamics.

\item Implicit large eddy simulation: A notable effort has been made to refine the SVV-based strategy for under-resolved simulations of high Reynolds fluid flows
  \cite{MOURA2016401,MOURA2015695}. This strategy has great potential to provide high fidelity iLES computations for a broad range of applications in hydrodynamics.

\item Fluid-structure interaction: As indicated in
  \cite{priolo1994,NME:NME1054,Bodard2006,SPRAGUE20062149,SPRAGUE2003149},
  spectral elements offer superior wave propagation capabilities,
  which has been used for simulations of fluid-structure interaction
  ({\it e.g.}\cite{Bodard2006,SPRAGUE20062149,VANLOON2007833}). Interaction subject to surface waves can be attractive

\item Cavitating flow: Recently, discontinuous spectral/{\it hp}
  element method has been implemented for the efficient and accurate
  simulation of multiphase flows \cite{Hoffmann_ICCFD}.  The
  compressible Navier-Stokes equations, coupled with a highly
  realistic equation of state, are solved by a discontinuous
  Galerkin spectral element method for flows with cavitation.  In {\it
    Nektar}++, the discontinuous Galerkin method and the flux
  reconstruction approach have been used to solve the compressible
  Navier-Stokes equations. This therefore provides a basis to apply the
  spectral/{\it hp} element method in applications of cavitating
  flows.
\item Free surface flows: As is well known free surface flows can be modelled as a depth average “shallow-water” approach or using ALE moving mesh techniques. The spectral/{\it hp} element method has been used to treat these two kinds of free surface flows \cite{RobertsonSherwin1999,LEEWING1990355,Sussman2003,FLD:FLD1006,Bouffanais2006}). For the second free surface problem, discontinuous spectral/{\it hp} element methods have a great potential in applications ({\it e.g.} \cite{Sussman2003,FLD:FLD1006,GROOSS20063406,MARCHANDISE2006780,NGUYEN20101}), especially when {\it h}/{\it p}-adaptivity is employed.
\end{enumerate}

%% file: acknowledgments.tex
\section{Acknowledgments}

The authors kindly thank the Executive Editor-in-Chief Prof. Liandi Zhou for  the invitation to contribute this review article and Dr. Wei Zhang for his contribution to conducting the tip-vortex simulation. H.X. and S.J.S would like to acknowledge support under EPSRC grant EP/L000407/1.